\documentclass[showpacs,twocolumn,preprintnumbers,amsmath,amsfonts,amssymb,floatfix,aps,superscriptaddress]{revtex4}

\usepackage{graphicx}
\usepackage{epsfig}
\usepackage[hang,nooneline]{subfigure}
\usepackage[normalem]{ulem}
\usepackage{color}
\usepackage{hyperref}

\graphicspath{{./fig-jpg/}{./fig-ps/}{./figures/}}

\newcounter{fig}

\definecolor{mygreen}{rgb}{0,0.6,0}

\begin{document}

\title{Planar and Radial Kinks in Nonlinear Klein-Gordon Models:\\
Existence, Stability and Dynamics}
\author{P. G. Kevrekidis}
\affiliation{Department of Mathematics and Statistics, University of Massachusetts
Amherst, Amherst, MA 01003-4515, USA}
\author{I. Danaila}
\email[Email: ]{ionut.danaila@univ-rouen.fr}
\affiliation{Laboratoire de Math{\'e}matiques Rapha{\"e}l Salem, Universit{\'e}
  de Rouen Normandie, 76801 Saint-{\'E}tienne-du-Rouvray,  France}
\author{J.-G. Caputo}
\email[Email: ]{caputo@insa-rouen.fr}
\affiliation{Laboratoire de Math{\'e}matiques Rapha{\"e}l Salem, Universit{\'e}
  de Rouen Normandie, 76801 Saint-{\'E}tienne-du-Rouvray,  France}
\affiliation{Laboratoire de Math{\'e}matiques, INSA de Rouen Normandie, France}
\author{R. Carretero-Gonz{\'a}lez}
\affiliation{Nonlinear Dynamical Systems
Group,\footnote{\texttt{URL}: http://nlds.sdsu.edu}
Computational Sciences Research Center, and
Department of Mathematics and Statistics,
San Diego State University, San Diego, California 92182-7720, USA}

\date{\today \,; to appear {\bf in Physical Review E}}

\begin{abstract}
We consider effectively one-dimensional
planar and radial kinks in two-dimensional nonlinear Klein-Gordon models
and focus on the sine-Gordon model and the $\phi^4$ variants thereof. 
We adapt an adiabatic invariant formulation recently developed for nonlinear 
Schr{\"o}dinger equations, and we study the transverse stability of these kinks.
This enables us to characterize one-dimensional planar kinks 
as solitonic filaments, whose stationary states and corresponding
spectral stability can be characterized not only in the
homogeneous case, but also in the presence of external potentials.
Beyond that, the full nonlinear (transverse) dynamics of
such filaments are described using the reduced, one-dimensional,
adiabatic invariant formulation.
For radial kinks, this approach confirms their azimuthal stability.
It also predicts the possibility of creating stationary and stable 
ring-like kinks.
In all cases we corroborate the results of our methodology with
full numerics on the original sine-Gordon and $\phi^4$ models.
\end{abstract}

\maketitle

\section{Introduction}

The study of the existence, transverse stability, and dynamics of
coherent structures that have an effective dimensionality
lower than that of the space in which they live,
is one that has a time-honored history in areas
such as nonlinear optics~\cite{agrawal} and atomic physics,
especially in connection with Bose-Einstein
condensates (BECs)~\cite{stringari,siambook}.
This is, among other reasons,
due to the remarkable observation and associated
analysis of the potential of coherent structures
to undergo transverse instability~\cite{kuzne,kidep} which leads
to the spontaneous formation of structures that are particular
to (and more robust within) the higher-dimensional setting,
such as vortices in two-dimensional (2D) settings~\cite{pismen}, and vortex lines and rings
in 3D settings~\cite{siambook}. It is worthwhile to note that
this type of instability, e.g., for prototypical structures such
as the so-called dark solitons (which are ubiquitous in both
nonlinear optics~\cite{Kivshar-LutherDavies} and atomic BECs~\cite{djf})
has been explored extensively at the experimental level too.
In particular, the production of vortices in the former
setting~\cite{tikh} and vortex rings in the latter~\cite{watching}
through this mechanism has been verified. This, in turn,
has made this a subject of persisting theoretical interest aimed
both towards analyzing and understanding such instabilities \cite{smirnov,hoefer,wang1,wang2},
as well as towards avoiding them \cite{us}.

As a related topic, it should be mentioned that higher dimensional
settings also enable the consideration of different geometric configurations,
e.g. ones with different curvature etc. In particular, with regard to
the kinklike dark soliton structures, naturally an extension of
such a 1D heteroclinic structure to 2D may involve a planar front
(like a 1D wall separating left from right or top from
bottom). However, it is also possible to form such structures
in a ring-like shape. The latter pattern, the so-called ring dark
soliton has again been explored both in optics~\cite{kivyang,rings,rings1}
and in BECs~\cite{rings2,rings3,rings4}. 
Extensions to even higher dimensions such as shells of either planar
or spherical
form have also been explored
in 3D; see, e.g.,~\cite{kivyang,carr,wenlong,hau} among many others.

The focus of the present work is to generalize
some of the ideas that have recently proved useful in analyzing
such structures in atomic BECs~\cite{wang1,wang2} to another
setting with a time-honored history, namely Klein-Gordon (KG)
equations. Some prototypical examples among these field theories
consist of the sine-Gordon (sG) equation, to which whole volumes
have been dedicated~\cite{sgbook}, as well as the $\phi^4$ model.
The latter is among the principal models for phase transitions
in statistical physics~\cite{parisi}, a toy model widely
used in ferroelectrics, polymeric chains, and nuclear
physics among many others~\cite{schonfeld}, but also a classical
(as well as quantum) field theory of particular interest in
its own right~\cite{belova}.
In these KG settings, which have been extensively explored
over the years (see e.g.~Ref.~\cite{kivsharmalomed} for a review),
the study of excitations such as, e.g., radial kinks has been
a topic of particular interest from early on. One can note
numerous attempts to explore the kink dynamics in two-
and even in three-dimensions~\cite{christiansen,geicke,bogolub},
as well as to develop equations of motion, e.g., for moving radial
kinks~\cite{samuelsen}, or to appreciate the rate of radiation of
shrinking radial kinks~\cite{malo}. Efforts along similar lines
both theoretically and numerically have been pursued for
the $\phi^4$ model~\cite{maloma,gleiser}. A recent summary
of the relevant earlier activity, and suggestion to utilize
radial kinks as sources of a fast breather (emerging from their
detrimental collision with boundaries) can be found in Ref.~\cite{caputo}.

Our aim in the present work is to adapt some of the earlier ideas
presented in the context of transverse instabilities in the
nonlinear Schr{\"o}dinger (NLS) settings in Ref.~\cite{wang1,wang2}
to the realm of the KG prototypical models (sG and $\phi^4$).
In previous works, various authors have described the evolution of radial 
kinks due to curvature effects \cite{christiansen,geicke,bogolub,samuelsen}.
However, these works only focused on the case of purely radial dynamics
where the kink remains perfectly circular and does not undergo any
transverse perturbation ---with the notable exception of Ref.~\cite{Boris},
which described the dynamics of elliptical solutions (pulsons) 
in the sG model.
In contrast, in the present work, formulating a Hamiltonian 
framework for KG models and adapting the solitonic filament method
will allow us to examine the existence and stability of longitudinal kinks
in two dimensions and, importantly, to formulate reduced PDEs
for their transverse evolution. Furthermore, this methodology
will also enable us to understand what role external (nonlinear) 
potentials can play in either stabilizing or destabilizing such kinks.
Such potentials are certainly possible in practical applications.
For instance, in Chap.~1 of Ref.~\cite{sgbook}, the presence of
potential terms in the form considered herein
has been connected to the presence of spatial
inhomogeneities in the context of Josephson junctions;
see also Ref.~\cite{mal27}.
Another example of this type is the so-called Josephson
window junction, leading to the ``dressing'' (width variation)
of the kinklike fluxon~\cite{jgc2}.
We will argue that not only are such
potentials interesting in their own right, but rather they will also 
serve to create an unprecedented example of a stable radially
symmetric kinklike structure in both prototypical KG
models. As an aside, we will unveil how the transverse instabilities
that are quite detrimental for kinklike (dark soliton) structures
in optics/BEC are absent in the KG settings considered.
Instead, the transverse undulations will be seen to be of a benign
oscillatory character. Overall, we believe that this perspective
will shed light on the (as we will call it) filamentary
dynamics of kinks  in higher dimensional
KG models, and it will open new directions for their stabilization
and practical use in applications, such as Josephson junction
arrays~\cite{sgbook}.

Our presentation is structured as follows.
In the next section, we detail the theoretical analysis of
the transverse dynamics of quasi-1D structures. We start by
recalling the instructive example of NLS from earlier works~\cite{wang1,wang2}
to which we later compare our KG case examples. We present
the theory for KG structures by first focusing on the 
case of ``standard'' rectilinear 1D kinks embedded within a 2D domain.
Then, we present the more involved case of radial kinks.
In Sec.~\ref{sec:numerics} we showcase the theoretical results
presented in Sec.~\ref{sec:theo} by comparing the stability
predictions and dynamics of our approach against the corresponding 
ones for the original KG models.
Finally, in the last section, we conclude by presenting a 
number of challenges for future work in this theme.

\section{Theoretical Analysis}
\label{sec:theo}

\subsection{A Preamble: the NLS Case}

We start our theoretical analysis by briefly revisiting the
planar NLS dark soliton in the 2D setting. The
model, in that case relevant to atomic Bose-Einstein condensates
as well as nonlinear optics~\cite{stringari,siambook,agrawal},
is the defocusing nonlinear Schr{\"o}dinger of the form:
\begin{eqnarray}
  i u_t =-\frac{1}{2} u_{xx} + |u|^2 u + V(x) u.
  \label{id_eq1}
\end{eqnarray}
When the potential is absent (i.e., $V(x)=0$), the energy of
the model is conserved and it has the functional form:
\begin{eqnarray}
  H_{\rm 1D}=\frac{1}{2}  \int_{-\infty}^{\infty} \left[|u_x|^2 +
  \left(|u|^2-\mu \right)^2\right] dx,
\label{id_eq2}
\end{eqnarray}
where the constant $\mu$ represents the chemical potential.
In this same $V(x)=0$ case, the prototypical
exact solitary waves of the model
are of the form~\cite{Kivshar-LutherDavies,djf}:
\begin{eqnarray}
u(x,t)=e^{-i \mu t} \left[
  \beta \tanh \left(\beta
(x-{\xi}) \right) + i v \right],
\label{id_eq3}
\end{eqnarray}
where $\beta=\sqrt{\mu - v^2}$ and the speed of the kink is $\dot{\xi}=v$.
The energy of such a configuration is [substituting Eq.~(\ref{id_eq3})
in Eq.~(\ref{id_eq2})] $H_{\rm 1D}=(4/3) (\mu-\dot{\xi}^2)^{3/2}$.
The fundamental idea of Ref.~\cite{konotop_pit} was to use this energy
as an adiabatic invariant (AI) even in the case in which there is a potential
with the modification that locally the chemical potential
$\mu$ becomes  $\mu-V(x)$ in the presence of such a term. From this
AI quantity:
\begin{eqnarray}
  H_{\rm 1D} = \frac{4}{3} \left(\mu - V({\xi}) -\dot{\xi}^2 \right)^{3/2},
  \label{id_eq4}
\end{eqnarray}
one can successfully infer the equation of motion of a 1D dark soliton as:
$\ddot{\xi}=- \frac{1}{2} V'({\xi})$.

Our interest is in generalizing this idea to higher dimensional
settings, extending the ideas of NLS to the KG class of models.
Thus, to complete our recap of the former~\cite{wang1,wang2},
we note that in the NLS case the 2D model
reads~\cite{agrawal,stringari,siambook}:
  \begin{eqnarray}
    i u_t =-\frac{1}{2} \left(u_{xx}+u_{yy}\right) + |u|^2 u + V(x) u,
    \label{id_eq5}
  \end{eqnarray}
The corresponding energy (for the $V(x)=0$ case) is also conserved in the form:
  \begin{eqnarray}
  H_{\rm 2D}=\frac{1}{2}  \iint_{-\infty}^{\infty} \left[ |u_x|^2 + |u_y|^2 +
    \left(|u|^2-\mu \right)^2 \right] dx\, dy.
  \label{id_eq6}
\end{eqnarray}
  Substituting now the expression of Eq.~(\ref{id_eq3}), but with
  the center $\xi$ being a function $\xi=\xi(y,t)$, we obtain,
  for the $V(x)\not=0$ case, the AI
  of the form:
 \begin{eqnarray}
  E= \frac{4}{3} \int_{-\infty}^{\infty} \left[ \left(1 + \frac{1}{2}{{\xi}}_y^2 \right)
  \left(\mu - V({\xi}) -{{\xi}}_t^2 \right)^{3/2} \right] dy.
  \label{id_eq7}
\end{eqnarray}
 Then, as explained in Ref.~\cite{wang2}, one can take the derivative $dE/dt=0$,
 taking advantage of the adiabatic invariance of this quantity and
 from that (and a suitable integration by parts), derive the effective
 equation of motion of a single dark soliton {\em filament} in a
 transverse modulated (potential) environment in the form:
 \begin{eqnarray}
  {{\xi}}_{tt} {\cal B} + \frac{1}{3} {{\xi}}_{yy} {\cal A} = {{\xi}}_y\, {{\xi}}_t\, {{\xi}}_{yt}
  -\frac{1}{2} V'({\xi}) \left( {\cal B} - {{\xi}}_y^2 \right),
\label{id_eq8}
\end{eqnarray}
where
${\cal A}=\mu - V({\xi}) - {{\xi}}_t^2$ and ${\cal B}= 1 + \frac{1}{2} {{{\xi}}_y^2}$.
This result encompasses [if $\xi=\xi(t)$] the 1D result as a special case;
its linearized, homogeneous (i.e., $V=0$) case retrieves the famous
transverse instability of Ref.~\cite{kuzne} (see also Ref.~\cite{kidep} for a
review). Namely, in that case we obtain:
\begin{eqnarray}
  {{\xi}}_{tt} + \frac{\mu}{3} {\xi}_{yy} =0,
  \label{id_eq8a}
\end{eqnarray}
which leads to a dispersion relation $\omega =\pm i \sqrt{\frac{\mu}{3}} k$,
between the frequency $\omega$ and the wavenumber $k$, indicating
instability.
Importantly, Eq.~(\ref{id_eq8}) also provides a reduced, effective
description of the 2D nonlinear
dynamics of the solitonic filament through the evolution of its
center as a function of the transverse ($y$) variable and time.
The aim of the present work is to present such a calculation for
the KG case and to appreciate its implications in connection and
in comparison with the NLS one.

\subsection{KG Planar Kinks}

We now turn to KG models which in general in their
prototypical 2D format read:
\begin{eqnarray}
  u_{tt}=\Delta u - (1 + V_{\rm ext}(x,y)) V'(u),
  \label{id_eq9}
\end{eqnarray}
where $V_{\rm ext}(x,y)$ is an external potential and $V(u)$ is the
intrinsic potential that defines the particular KG model at hand:
$V(u)= 1-\cos(u)$ for the sine-Gordon model and $V (u)= (u^2 -1)^2/2$ 
for the $\phi^4$ model.
In fact, the most canonical form is that of $V_{\rm ext}(x,y)=0$ in homogeneous
space, but our aim here is to use the above adiabatic invariant phenomenology
to appreciate the effect of such spatial inhomogeneities on the existence
and stability of solitary structures of the model.

More specifically, such a model has a conserved energy of the form:
\begin{eqnarray}
  H_{\rm 2D}&=& \iint_{-\infty}^\infty  \left[\frac{1}{2} \left(u_t^2 + u_x^2 + u_y^2 \right)
  \right.
\nonumber
\\[1.0ex]
&&+ \left. \phantom{\frac{1}{2}}(1 + V_{\rm ext}(x,y)) V(u) \right] dx\, dy.
  \label{id_eq10}
\end{eqnarray}
Our starting point will be to assume a quasi-1D kink in the present
section. Next, we will consider the less straightforward case of
a radially symmetric kink. Thus, we use the ansatz of
the form
\begin{eqnarray}
  u(x,y,t)=f(x-X(y,t)),
  \label{id_eq11}
\end{eqnarray}
describing a kink of shape $f$ and position $X(y,t)$ modulated in time and,
importantly, along the $y$-direction.
In a model such as, e.g., the sine-Gordon (sG) model with
$V(u)=1-\cos(u)$, the kink is
$f(s)=4 \arctan(\exp(s))$, while for the
case of the $\phi^4$ model with $V(u)=(u^2-1)^2/2$, we have
$f(s)=\tanh(s)$.
It is important to highlight in both cases that in
the present work,
the relativistic effects, discussed, e.g., in Ref.~\cite{caputo} have been
neglected. Furthermore, in using this functional form, we
are assuming that we have a quasi-1D kink that is allowed
to be transversely modulated by the presence of the external
potential. The theory that we will develop is best
suited to the case when the $V_{\rm ext}$ is longitudinal in nature,
namely $V_{\rm ext}=V_{\rm ext}(x)$, but bears no dependence on $y$.
In fact, our numerical computations will suggest how to potentially
generalize things in the most general case, but the latter is outside
the scope of the present work. In such a case of a longitudinally
dependent external potential, we can perform the integration over
$x$ within Eq.~(\ref{id_eq10}) [upon substitution of Eq.~(\ref{id_eq11})],
in order to obtain the following expression:
\begin{eqnarray}
  E=\int_{-\infty}^\infty  dy \left[ \frac{1}{2} M \left( X_t^2 + X_y^2 \right)
  + E_{\rm 1D}^{\rm 1K} + P(X) \right].
  \label{id_eq12}
\end{eqnarray}
The first two terms in this energy stem from the kinetic energy
and also from the ``filamentary'' dependence of the kink
on the transverse variable [the $u_y$ term in Eq.~(\ref{id_eq10})].
Both are multiplied by the effective mass of the kink along the
longitudinal direction defined as $M=\int f'(s)^2 ds$.
For our KG models of choice $M=8$ for the sG and $M=4/3$ for
the $\phi^4$ cases, respectively.
The third
term in the energy stems from the combination of the $u_x$
term in the original energy and the unperturbed contribution in the
potential energy which combine to yield the energy of the single
kink (1K) in 1D (hence the superscript and subscript). This
quantity turns out to be $E_{\rm 1D}^{\rm 1K}=M$ in the cases under consideration.
Importantly notice that in the infinite domain limit this quantity
will yield an irrelevant divergence (as it is simply constant) due
to its proportionality to the domain size upon integration.
In the examples of interest in our case, all computations will be performed
in a {\em finite} computational domain, extending from $-l_y$ to $l_y$
i.e., the total length $L_y=2 l_y$.
As a result, this term is finite and remains bounded and thus
will not contribute towards the filament's dynamics. Finally,
\begin{eqnarray}
  P(X)&=&\int_{-\infty}^{\infty} V_{\rm ext}(x) G(x-X) \,dx
\nonumber
\\
  \Rightarrow P'(X)&=&\int_{-\infty}^{\infty} V_{\rm ext}'(x) G(x-X) \,dx,
  \label{id_eq13}
\end{eqnarray}
where in the last equality we have taken into account the suitable decay
of the solution (typically exponential for the profiles considered)
and $G(s)=V(u(s))$. 

Now, using $dE/dt=0$, i.e., the adiabatic invariance of the energy, one
obtains an equation of motion for the solitonic filament in the
transverse ``landscape''. More specifically, we will have
\begin{eqnarray}
  X_{tt}=X_{yy}-\frac{1}{M} P'(X).
  \label{id_eq14}
\end{eqnarray}
This principal result already has a number of interesting ramifications.
%
Firstly, it is relevant to connect the adiabatic dynamics of a planar kink
in 2D with that of a 1D kink. Assuming that $X$ does not depend on $y$,
Eq.~(\ref{id_eq14}) reduces to the well-known Newton-like equation for a 1D kink
in the presence of an external potential (see for example Ref.~\cite{rfftp88}
in the Josephson junction context).
Secondly, in the absence
of an external potential, i.e., for $V_{\rm ext}=0$, one should
get hyperbolic dynamics, i.e., a wave like undulation of the
structure in the transverse direction. In the case of a finite
domain, as in our present scenario, the associated wavenumbers
of such undulation are $\omega=k \pi/L_y$, for integer $k$.
Hence, we should expect to observe such modes in the linearization
around a kink in the 2D setting; it is perhaps also remarkable
that this conclusion will hold true {\em irrespectively} of the
setting [and the particular potential $V(u)$]. It is also worthwhile
to compare this behavior to the NLS setting where the dark
solitonic kinklike structure is transversely unstable and
each one of the modes of such undulation is associated with a
transverse instability. That is to say, the two models behave
oppositely as regards the stability of transverse undulations
affecting their kinks. 
Namely, planar dark solitons for the defocusing 2D NLS 
Eq.~(\ref{id_eq1}) are unstable to transverse perturbations.
However, planar bright solitons for the focusing NLS
[same as in Eq.~(\ref{id_eq1}) with $-|u|^2u$]
exhibit collapse and are immune to transverse instabilities.
Along this train of thought, the KG model can be reduced,
via multiple scale analysis~\cite{agrawal}, to the focusing NLS. 
This connection helps to explain why KG planar kinks do not exhibit
transverse instabilities. 
%
%

Generalizing away from the homogeneous $V_{\rm ext}=0$ case, 
the methodology prescribed above allows for the possibility of engineering
scenarios (through appropriate choices of the external potential) 
leading to particular dynamics of the kink. In particular, one can engineer
whether a particular steady state configuration is stable or unstable depending on
the nature of the external potential. 
In fact, one can use Eq.~(\ref{id_eq13})
to engineer the external potential to achieve any desired form of $P'(X)$.
For instance, using the even nature (in the examples of interest) of
$G(x-X)$, we can rewrite:
\begin{eqnarray}
  P'(X)=V_{\rm ext}' \ast G
  \Rightarrow \hat{P'}=\hat{V_{\rm ext}'} \hat{G}
  \Rightarrow V_{\rm ext}'=\widehat{\left(\hat{P'}/\hat{G}\right)}.
  \nonumber
\end{eqnarray}
Here, $\ast$ has been used to denote convolution, while the
hat symbol has
been used to denote the Fourier transform (and
also its inverse). The final result suggests that given the
desired $P$ or $P'$, and for a particular model (meaning, given
$G$), one can reverse engineer the potential needed to
induce this ``force'' term $P'$ from Eq.~(\ref{id_eq13}).

To illustrate the kink dynamics described by our effective
filament Eq.~(\ref{id_eq13}), we consider
a simple and generic potential $V_{\rm ext}(x)=A\, {\rm sech}^2(x)$
that represents a localized barrier for $A>0$ or a localized well for $A<0$.
For such a potential,
%
%
$P'$ can be computed explicitly from Eq.~(\ref{id_eq13}) 
and for the two models of interest it reads:
\begin{eqnarray}
  P'(X)=-4 A\, {\rm csch}^4(X) \left( (2 + C) 2 X- 3 S\right),
  \label{id_eq16}
\end{eqnarray}
for the sG case, while it is:
\begin{eqnarray}
  P'(X)=-\frac{A}{3} {\rm csch}^6(X) \left(T -36 X -24 X C + 28 S \right),~
  \label{id_eq17}
\end{eqnarray}
for the $\phi^4$ model,
where $C\equiv\cosh (2 X)$, $S\equiv\sinh(2 X)$, and $T\equiv\sinh(4 X)$.
Importantly, in such a setting, in order to describe the
dynamics in the vicinity of the equilibrium $X=0$, one can
linearize around such a state. Then, a direct Taylor expansion,
taking into account once again the transverse extent of the domain
from $-l_y$ to $l_y$ (remember $L_y=2l_y$) yields 
the following linearization frequencies:
\begin{eqnarray}
  \omega= \pm \sqrt{-\frac{32 A}{15 M} + \left(\frac{k \pi}{L_y}\right)^2}
  \label{id_eq18}
\end{eqnarray}
for the sG case, while the corresponding prediction is:
\begin{eqnarray}
   \omega= \pm \sqrt{-\frac{64 A}{105 M} + \left(\frac{k \pi}{L_y}\right)^2}
  \label{id_eq19}
\end{eqnarray}
for the $\phi^4$ model.
Importantly, these analytical predictions give us immediate insight
into the modes that can potentially induce instabilities.
The most unstable among them is, naturally, the $k=0$ mode (since
we indicated that higher undulational modes are generally more robust).
Expressions (\ref{id_eq18}) and (\ref{id_eq19}) illustrate that 
if $A>0$, i.e.~the local potential acts as a barrier, then  the $k=0$ mode 
has a corresponding imaginary eigenfrequency so the kink is unstable, 
as expected from the effective Newton-like 1D picture. If $A<0$ the potential
corresponds to a local well, and thus, all the eigenfrequencies 
are real and the kink is spectrally stable.
%

\subsection{Radial KG Kinks}
\label{sec:radial}

Having considered the simpler case of planar KG kinks, it is relevant
now to extend considerations to the case of radial kinks which have
a more elaborate phenomenology. We will not present the radial NLS
case (as we are now quite familiar with the method),
but simply note that it has been elaborated in Ref.~\cite{wang1}
and yields the following conclusions. The curvature pushes the
defocusing NLS dark soliton outward, if it is started in a radial
configuration. Unless an external potential is imposed, it is thus
not possible for an equilibrium radial  dark soliton configuration to exist.
In the case of Bose-Einstein condensates~\cite{stringari,siambook},
the prototypical scenario involves trapping of the atoms through an
external parabolic trap which indeed has a confining, typically radial
effect that, in turn counters the role of curvature, creating the
possibility of a stationary ring dark soliton (see details in Ref.~\cite{djf}).
Nevertheless, this configuration is unstable to transverse
modulations and the adiabatic invariant
theory of Ref.~\cite{wang1} enabled
a systematic asymptotic capturing of these azimuthally growing modes.

Our aim here is to explore the analogous dynamics in the case of
sG and $\phi^4$ models which in their own right have a time-honored
history of explorations discussed in some detail in the Introduction.
Indeed, as summarized in the recent work of Ref.~\cite{caputo},
the effect of curvature in these KG models is exactly the opposite
of that of NLS. Namely, the curvature pushes an initial kink
centered at $r=R_0$ inward, forcing it to eventually collapse
at $r=0$ for an implosion that subsequently ejects outwards
fast breathers at least in the 2D problem (in 3D
the phenomenology can be more elaborate).

\begin{figure}[tb]
\begin{center}
\includegraphics[width=7cm]{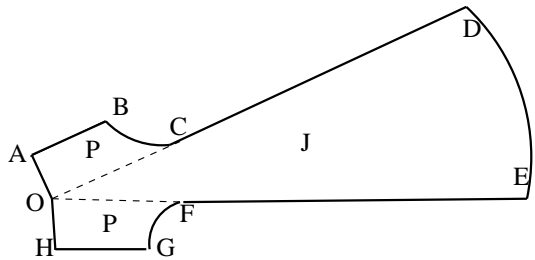}
\caption{Top view of a Josephson junction corresponding to the
2D sG Hamiltonian of Eq.~(\ref{id_eq20}). The region OCDEFO is
the junction area where the oxide layer is thin, enabling
Josephson coupling. In the two surrounding passive regions ABCOA
and OFGHO, labeled P, the oxide layer is thicker so that
no tunneling is present.}
\label{id_fig0}
\end{center}
\end{figure}

We start again from the energy of the KG models, this time written
in polar coordinates:
\begin{eqnarray}
  H_{\rm 2D}&=&\iint \left[\frac{1}{2} \left( u_t^2 + u_r^2 + \frac{1}{r^2} u_{\theta}^2
  \right) \right.
\nonumber
\\
&&+ \left. \phantom{\frac{1}{2}} \left(1+V_{\rm ext}(r,\theta)\right) V(u) \right] r\, dr\, d\theta, 
  \label{id_eq20}
\end{eqnarray}
where $0\le r$ and $0 \le \theta \le \theta_0$. The boundary
conditions at $\theta=0$ and $\theta=\theta_0$ are homogeneous Neumann,
$u_\theta =0$.
A physical realization of the Hamiltonian Eq.~(\ref{id_eq20}) 
with axial symmetry
is the composite Josephson junction shown
from the top in Fig.~\ref{id_fig0}.
A Josephson junction is a structure composed
of two superconductors separated by a thin film ($\sim 10$\AA)
enabling tunneling between the films; the term $V(u)=1 -\cos(u)$ is
due to the tunneling. The device shown in Fig.~\ref{id_fig0}
has two adjoining passive
regions where no tunneling is present, and this modulates $V(u)$
which can then be described by $V(u) V_{\rm ext}(r)$. This effect was studied
in detail, e.g., in Ref.~\cite{vavalis}.
In the hereby proposed device, the nonlinearity is reduced
close to $r=0$ and restored to its value away from $0$.
In what follows, we focus on the case in which $\theta_0=2\pi$. 
Nonetheless, it is important to stress that the theory and 
results presented below are equally applicable to the case in which
$\theta_0<2\pi$ by simply adjusting the relevant $\theta$-integrals
to be from $\theta=0$ to $\theta=\theta_0$.
%
If $\theta_0=2\pi$, instead of the formulation of Fig.~\ref{id_fig0}, 
one could consider a circular Josephson junction that includes an
inner and an outer disk with different thicknesses so as to change
the local properties of the device from the center outward. In this manner,
by controlling the thickness of these disks, it is conceivable to 
design different desirable two-dimensional external potentials.

Now, we seek a radial solution of the form:
\begin{eqnarray}
u(r,\theta,t)=f(r-R(\theta,t)),
\end{eqnarray}
that is to say a kink
centered at $R(\theta,t)$, i.e., a filament (topologically
equivalent to a circle) that potentially undulates
along the transverse (azimuthal) direction. The relevant substitution
in Eq.~(\ref{id_eq20}) yields:
\begin{eqnarray}
  E=\int_0^{2\pi} \left[ \frac{1}{2} M R R_t^2 + E_{\rm 1D}^{\rm 1K} R + \frac{1}{2}
    \frac{M}{R} R_\theta^2 + P(R) \right] d\theta.
    \label{id_eq21}
\end{eqnarray}
It is important to explain the terms in this expression, as well
as the nontrivial assumptions that they implicitly harbor (as these
will be responsible for the limitations of the theory in what follows).

The first term naturally stems from the kinetic energy. However, the
exact form of this term would be as follows, using $s=r-R$:
\begin{eqnarray}
  E_K&=&\frac{1}{2} \int_0^{2\pi}\!\! d\theta \int_{0}^{\infty} r\, dr\, (f'(r-R))^2 R_t^2
  \label{id_eq22}
\\[2.0ex]
\nonumber
  &=& \frac{R_t^2}{2} \int_0^{2\pi}\!\! d \theta \int_{-R}^{\infty} ds (s+R) (f'(s))^2\\ \nonumber
 & \approx&  \int_0^{2\pi}  \frac{1}{2} M R R_t^2\,\, d \theta.
\end{eqnarray}
As mentioned above, the integrations over $\theta$ are from $0$ to $2 \pi$. 
More significantly in the last step the approximations involved are
the following:
\begin{itemize}
\item  $(f')^2$ is an even function of its argument. This is fairly
  standard and is adhered too in the models of interest.
\item $\int_{-R}^{\infty} s (f'(s))^2 ds \approx 0$. This is implicitly
  assuming that the kink dies off sufficiently fast for an $R$ sufficiently
  large so that the exponentially small corrections of the integral
  from $-\infty$ to $-R$ will be negligible. This assumption will
  be very good for $R$ large and will obviously falter as $R \rightarrow 0$.
\item Similarly, for the nonzero contribution, again up to exponentially
  small corrections in $R$, it is assumed that $\int_{-R}^{\infty}
  f'(s)^2 ds \approx M$.
\end{itemize}
All of the terms that appear in Eq.~(\ref{id_eq22}) bear similar
assumptions. The curvature term involving $R_{\theta}^2$
was derived by
neglecting
higher order curvature driven terms, such as the ones arising due
to the Taylor expansion of $1/r$ around the kink center $s=0$ (equivalently
$r=R$).

Now let us make a remark before we move on to the
general discussion. Assume a purely radial dynamics, in the absence
of an external drive. Then, the azimuthal term does not contribute
and the integration over $\theta$ yields a factor of $2 \pi$.
Thus, the energy amounts to:
\begin{eqnarray}
  \frac{E}{2 \pi}= M  \left( \frac{1}{2} R R_t^2 + R \right).
  \label{id_eq23}
\end{eqnarray}
As it should, this expression is identical to the small speed limit of its relativistic
analogue  explored in Ref.~\cite{caputo} (see also the discussion therein
for earlier related works). The latter is given by $E=M R/\sqrt{1-R_t^2}$.
The work of Ref.~\cite{samuelsen} was apparently the first one to use
the constancy of this energy and to set it equal to its initial value
and integrate the result to find that for initially stationary kinks
(in the sG model), the trajectory of the inward moving kink is
cosinusoidal according to $R=R_0 \cos(t/R_0)$, which was found to be in
excellent agreement with numerical results. This prediction is valid
in the 2D case, while in the 3D case $R$ is replaced by $R^2$
leading to a cnoidal inward dynamics of the radial kink.
Notice that in our case, not having a priori assumed anything about the
potentially relativistic nature (and associated Lorentz invariance)
of the kink, we retrieve the classical limit thereof of small
speeds $R_t \ll 1$.

Now, let us move forward on the basis of the above assumptions.
What Eq.~(\ref{id_eq23}), and by extension Eq.~(\ref{id_eq21}),
implies is that (upon dividing by $R$ and up to a constant), the
effective radial dynamics involves a curvature induced effective
potential proportional to $1/R$, which is an attractive one toward the origin
and hence leads the kink to ``collapse'' within finite time
to $R=0$. The key question is whether
we can use an external effective potential $P(R)$, based on
the term $V_{\rm ext}(r)$ in the equations of motion that
will ``hold'' the kink up against such an inward motion and
eventually produce an effective radial equilibrium. At the same
time, it is of interest to examine what the fate of the transverse
perturbations is. In the NLS realm of ring dark solitons, they
cause transverse instabilities giving rise to vortices, while
here they are expected (on the basis
of the calculations of the previous subsection) to be benign.
Nevertheless, it is important to establish this quantitatively.

Differentiating the adiabatic invariant of the energy, we obtain
that
\begin{eqnarray}
\frac{dE}{dt}&=&\int_0^{2\pi} d\theta
\left[ \frac{M}{2} R_t^3 + M R R_t R_{tt}
- \frac{M}{2 R^2} R_t R_\theta^2 
\right.
\nonumber
\\
&&+ \left. \phantom{\frac{1}{2}}
 \frac{M}{R} R_\theta
R_{\theta t} + E_{\rm 1D}^{\rm 1K} R_t + P'(R) R_t \right].
\label{id_eq24}
\end{eqnarray}
Finally, from Eq.~(\ref{id_eq24}), one can obtain the equation of motion
(upon an integration by parts the fourth term in the integrand, and remembering that this quantity
should identically vanish for all choices of $R$)
\begin{eqnarray}
  M R R_{tt} + \frac{M}{2} R_t^2 + \frac{M}{2 R^2} R_{\theta}^2
  - \frac{M}{R} R_{\theta \theta}=-E_{\rm 1D}^{\rm 1K} - P'(R).
  \nonumber
\end{eqnarray}
Thus, in order for a steady state to exist, the
effective external force $-P'(R)/R$ needs to balance the
influence of the curvature $E_{\rm 1D}^{\rm 1K}/R$.
Namely, the equilibrium radius position $R_0$ can be approximated
by solving the equation
\begin{eqnarray}
\label{eq:R0}
P'(R_0)+E_{\rm 1D}^{\rm 1K}=0.
\end{eqnarray}
One can take this calculation one step further, assuming that
such a fixed point, say $R_0$, exists. In particular, we
linearize around $R_0$ as follows $R=R_0 + \epsilon R_1(t) e^{i n \theta}$.
Then, we obtain, upon renaming $Q(R)=P'(R)/R$ (and
recalling that $E_{\rm 1D}^{\rm 1K}=M$ for our models):
\begin{eqnarray}
  M \ddot{R}_1=\frac{M}{R_0^2} \left(1-n^2 \right) R_1 -Q'(R_0) R_1.
  \label{id_eq26}
\end{eqnarray}
Thus, using that $R_1(t) \sim e^{i \omega t}$, we obtain the final
expression:
\begin{eqnarray}
  \omega^2 = \frac{1}{R_0^2} (n^2-1) + \frac{1}{M} Q'(R_0).
  \label{id_eq27}
\end{eqnarray}

Now, we can make some comments on this result. First off, the
role of the transverse degrees of freedom is again a stabilizing
one. Clearly, the higher $n$ is, the higher is the relevant
eigenfrequency, and all the eigenvalues above a critical one
($n_{\rm cr}=\left[\sqrt{1+{R_0^2 Q'(R_0)}/{M}}\right]$ where $[\cdot]$
denotes the integer part), will by
necessity be stable. Nevertheless, whether the solution is
fully stable hinges critically on the contribution of
$V_{\rm ext}$. In particular, the most unstable eigenmode
is again the $n=0$ one, hence the stability of the
full structure is guaranteed, provided
\begin{eqnarray}
  Q'(R_0) \geq \frac{M}{R_0^2}.
  \label{id_eq28}
\end{eqnarray}
To offer perhaps a concrete example, forgetting temporarily
our assumption of large $R$,
one can envision Taylor expanding
\begin{eqnarray}
  V_{\rm ext}(r)&=&V_{\rm ext}(s+R)
  \label{id_eq29}
\\
\nonumber
   &=&V_{\rm ext}(s) + R V_{\rm ext}'(s) + \frac{R^2}{2}
  V_{\rm ext}''(s) + \dots
\end{eqnarray}
Then dubbing $V(u(r-R))=G(s)$, we can express
\begin{eqnarray}
  P(R)={\cal A} R + {\cal B} R^2 + ...
  \label{id_eq30}
\end{eqnarray}
where ${\cal A}=\int ds (V_{\rm ext}(s) + s V'_{\rm ext}(s)) G(s)$,
while ${\cal B}=\int ds (2 V_{\rm ext}'(s) + s V_{\rm ext}''(s)) G(s)$.
If we neglect higher than quadratic terms and dub $\tilde{{\cal A}}=
{\cal A}+ E_{\rm 1D}^{\rm 1K}$, then the equilibrium radius is given by $R_0=-\tilde{{\cal A}}/{\cal B}$
(and will exist only if $\tilde{{\cal A}} {\cal B} <0$), while the linearization
frequencies in this case will be
$\omega^2=(1/R_0^2) (n^2- {\tilde{{\cal A}}}/{M})$. Thus, for instance,
in such a setting one would need $\tilde{{\cal A}}<0$, ${\cal B}>0$ in order
to ensure stability. 

The above theory has important consequences. First, it
ensures the stability of transverse undulations of a radial kink
as observed in previous numerical simulations.
The theory also provides a set of guidelines to
ensure that the inward effect of curvature is balanced
by the outward effect, due to the external potential,
so that an equilibrium may be possible.
We will see below that this is indeed the case, so that a stable bound
state radial kink can be found when introducing a local potential well
at the origin. 
%

\section{Numerical Results}
\label{sec:numerics}

\subsection{Planar KG Kinks}

\begin{figure}[tb]
\begin{center}
\includegraphics[width=\columnwidth]{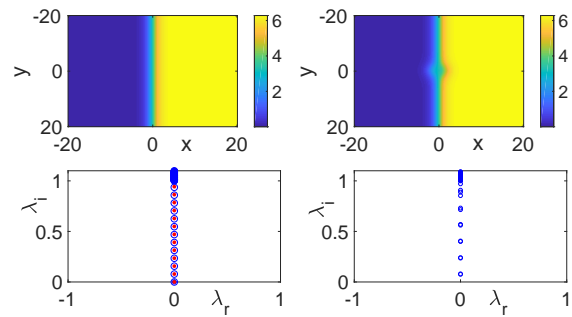}
\caption{(Color online)
The top panels show examples of the sG
  kinks with a radial external potential of the form
  $V_{\rm ext}(r)=A\, {\rm sech}(r)$. The left panels are
  for $A=0$ (i.e., in the absence of the potential for
  a ``standard'' 2D sG model), while the right panels
  are for $A=-4$. The bottom panels show the eigenvalues
  associated with the kinks in the linearized stability analysis.
  In the bottom left panel,
  the numerically computed eigenvalues from the full system (\ref{id_eq31})
  (blue circles) are found in excellent agreement with the theory (red stars), 
  see Eq.~(\ref{id_eq18}) with $A=0$, namely $\lambda=ik\pi/L_y$.
}
\label{id_fig1}
\end{center}
\end{figure}

\begin{figure}[tb]
\begin{center}
\includegraphics[width=\columnwidth]{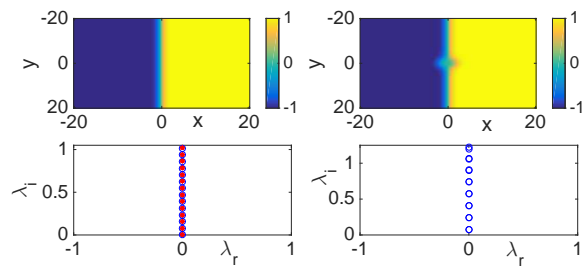}
\caption{(Color online)
Exactly the same panels as with Fig.~\ref{id_fig1}, and with
  exactly the same potential (and for the cases $A=0$ (left) and $A=-4$
  (right)), but now for the $\phi^4$ model, to illustrate the generic nature
  of our findings.}
\label{id_fig2}
\end{center}
\end{figure}

Turning now to the numerical examination of the results, we
start with the case of the planar KG kinks in both
the sG and the $\phi^4$ models.
We first consider a radially localized potential of the form
$V_{\rm ext}(r)=A\, {\rm sech}(r)$ ($r^2=x^2+y^2$).
Steady states, $u_0(x,y)$, of Eq.~(\ref{id_eq9}) are first found using standard
continuation techniques by discretizing space using a second order
finite difference scheme.
In Fig.~\ref{id_fig1}
and~\ref{id_fig2}, respectively, the case examples
in the absence of external potential (i.e., $A=0$)
have been considered in the corresponding left
panels of the figures. This is the scenario that, from a
solitonic filament perspective corresponds to the case
of $P'(X)=0$. Here, we will examine not only the existence
of an associated stationary kink, but also that of its spectral
stability. In particular, we perturb the kink solution $u_0(x,y)$
of Eq.~(\ref{id_eq9}) according to
\begin{eqnarray}
\label{eq:perturb}
 u(x,y,t)=u_0(x,y) + \epsilon e^{\lambda t} w(x,y),
\end{eqnarray}
and we solve the spectral problem
associated with the linearization in the form:
\begin{eqnarray}
  \lambda^2 w=\Delta w - \left(1+ V_{\rm ext}(x,y)\right) V''(u_0) w.
  \label{id_eq31}
\end{eqnarray}
Here, $\epsilon$ is a formal small parameter, $\lambda$ are the eigenvalues
of the linearization (if imaginary, suggesting the stable, oscillatory
nature of an eigenmode, while if real, indicating its instability),
and $w$ is the corresponding eigenvectors.
Note that an eigenvalue $\lambda$ corresponds to an eigenfrequency 
$i\omega=\lambda$.

As illustrated in the previous section, for $A=0$, {\em irrespectively of the model}, the kink
is supposed to satisfy a wave equation. This implies that
linearization around the equilibrium kink filament will
solely involve frequencies of oscillation according
to $\omega=k \pi/L_y$. Of course, this is in addition to the
continuous spectrum of the problem (i.e., the linearization
around the uniform states, on top of which the kinks exists),
which consists of the frequencies $\omega \in \pm [1,\infty)$ 
for the sG model.

We observe in Fig.~\ref{id_fig1} that both the $A=0$ and $A=-4$ cases 
are dynamically stable with all of the corresponding eigenvalues on 
the imaginary axis. 
Furthermore, as evidenced in the bottom-left panel of Fig.~\ref{id_fig1},
our approach captures the {\em entire} spectrum of the linearization
around the coherent structure. Indeed, for $A=0$, {\em all} the modes
predicted by the theory are identified in the stability analysis
in excellent agreement between the two. In the $\phi^4$ case, see Fig.~\ref{id_fig2},
the picture is rendered somewhat more complex due to the
presence of the internal mode at $\omega= \pm \sqrt{3}$.
Nevertheless, in addition to the continuous spectrum
which in this case is for $\omega \in [2, \infty)$, we can notice
the excellent matching between theory and numerics for
the undulational modes of the
unperturbed kink. To offer some perspective on the fact
that such planar kinks can even be perturbed
---without being destabilized--- in the
non-longitudinal direction, we have included a
radial potential of the form $V_{\rm ext}(r)=-4\, {\rm sech}(r)$.
As depicted in Figs.~\ref{id_fig1} and \ref{id_fig2},
the kink {\em remains} stable after the addition of this
radial potential,
solely incurring a small deformation in the
vicinity of its center in the area of action of the heterogeneous
external potential.

\begin{figure}[tb]
\begin{center}
\includegraphics[width=\columnwidth]{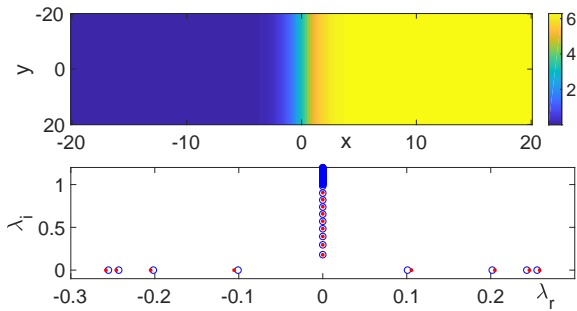}
\caption{(Color online)
An example of the sG model with a
  longitudinal potential $V_{\rm ext}(x)=A\, {\rm sech}^2(x)$ that
  conforms to the kink symmetry. Importantly, this potential
  renders the kink immediately unstable. The case example
  of $A=0.25$ is shown in the top panel for the kink profile,
  while the bottom shows the corresponding eigenvalues as
  computed numerically (blue circles) and as calculated by the
  analytical theory (red stars), see Eq.~(\ref{id_eq18}).
}
\label{id_fig3}
\end{center}
\end{figure}

\begin{figure}[tb]
\begin{center}
\includegraphics[width=\columnwidth]{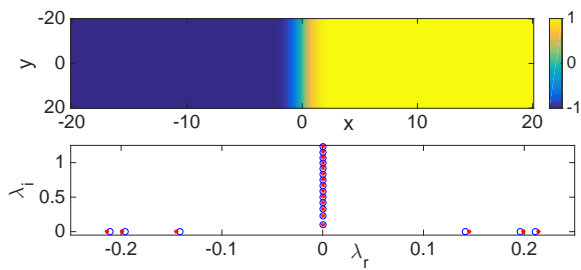}
\caption{(Color online)
Exactly the same panels as with Fig.~\ref{id_fig3}, and with
exactly the same potential but now for the $\phi^4$ model, and specifically
for $A=0.1$. 
Theoretical values (red stars) are estimated from Eq.~(\ref{id_eq19}).
}
\label{id_fig4}
\end{center}
\end{figure}

Now, to touch base with the theory developed in the previous
section, we examine a case in which the potential is along the
longitudinal direction and is of the form $V_{\rm ext}(x)=A\, {\rm sech}^2(x)$.
Notice that now Eqs.~(\ref{id_eq16}) and (\ref{id_eq17}),
respectively, apply for the sG and $\phi^4$ models. Furthermore,
more practically related to the computations in Figs.~\ref{id_fig3}
(for sG) and~\ref{id_fig4} (for $\phi^4$), Eqs.~(\ref{id_eq18})
and~(\ref{id_eq19}) are applicable. From these, we see immediately
that the selection of a value of $A>0$ will lead to instability
for all values of $A \neq 0$, while a choice of $A<0$ will lead
to spectral stability. Indeed, we consider in the cases of
Figs.~\ref{id_fig3} and~\ref{id_fig4}, particular examples of instability,
in order to test the validity of the theory and its ability to capture
both the unstable and also the stable modes of the solitonic filament.
In both cases, we see that the eigenvalue
predictions of Eqs.~(\ref{id_eq18})
and~(\ref{id_eq19}) are generally in very good agreement with the theory
(although the theory slightly overestimates the corresponding 
growth rates).

\begin{figure}[tb]
\begin{center}
\includegraphics[height=4.25cm]{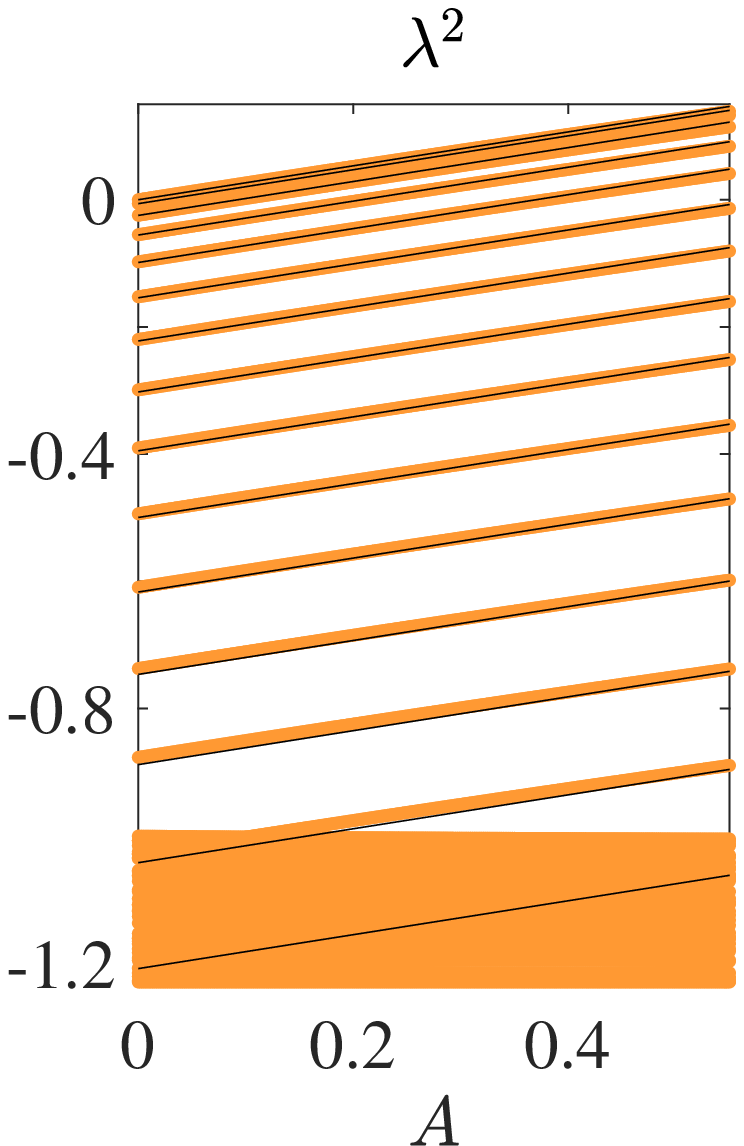}
~
\includegraphics[height=4.20cm]{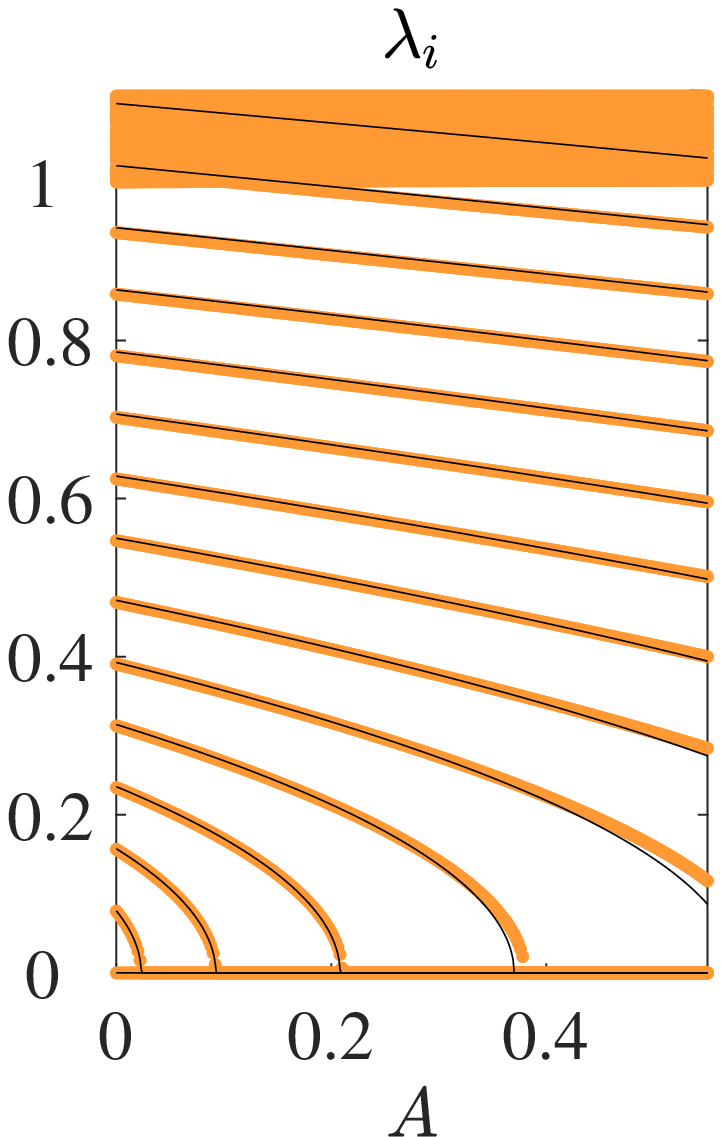}
~
\includegraphics[height=4.20cm]{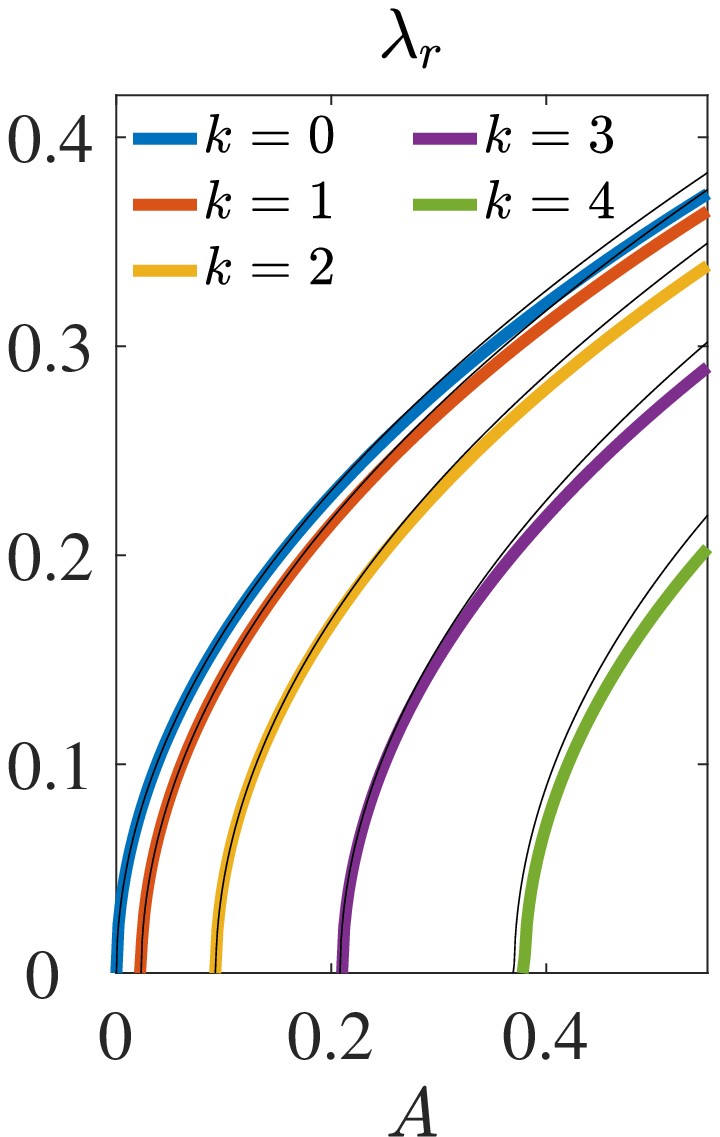}
\caption{(Color online)
Stability spectra for the sG model as a function of the 
amplitude of the longitudinal potential $V_{\rm ext}(x)=A\, {\rm sech}^2(x)$.
The left, middle, and right panels depict, respectively, $\lambda^2$,
$\lambda_i={\rm Im}(\lambda)$, and $\lambda_r={\rm Re}(\lambda)$.
The thick colored lines corresponds to the numerically computed
spectra and the thin black lines to our theoretical predictions
given by Eq.~(\ref{id_eq18}).
Note that for the unstable modes (see the right panel) the different
$k$-modes have been identified.
}
\label{fig:evals_sG}
\end{center}
\end{figure}

\begin{figure}[tb]
\begin{center}
\includegraphics[height=4.15cm]{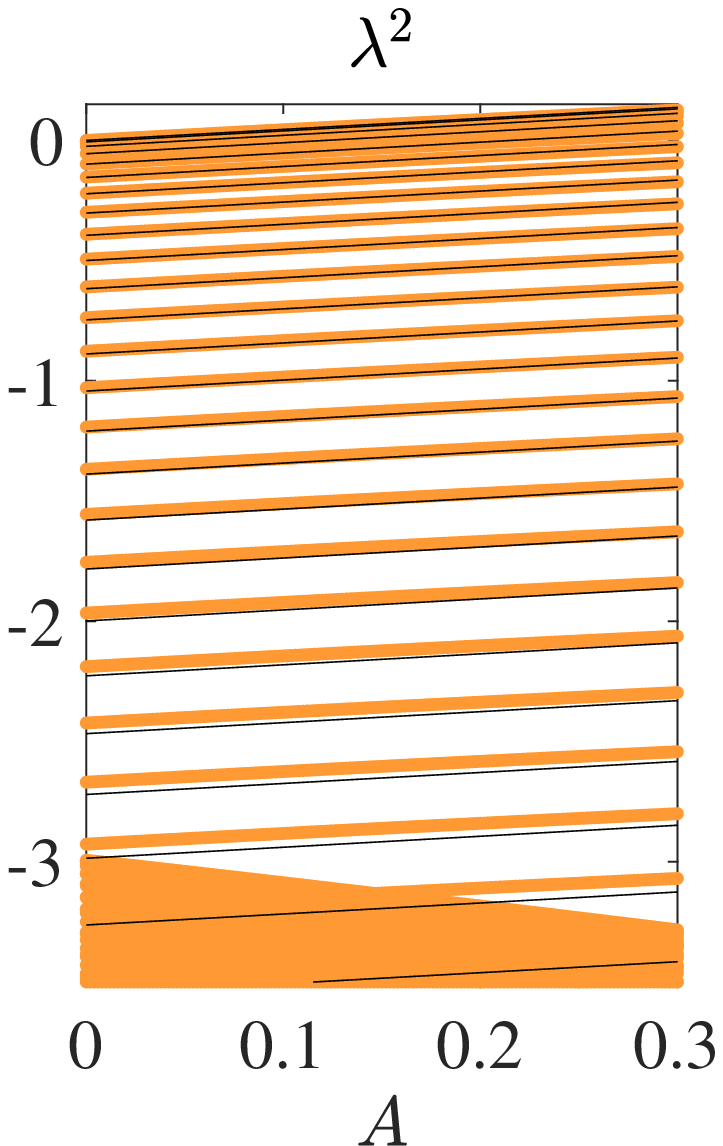}
\includegraphics[height=4.10cm]{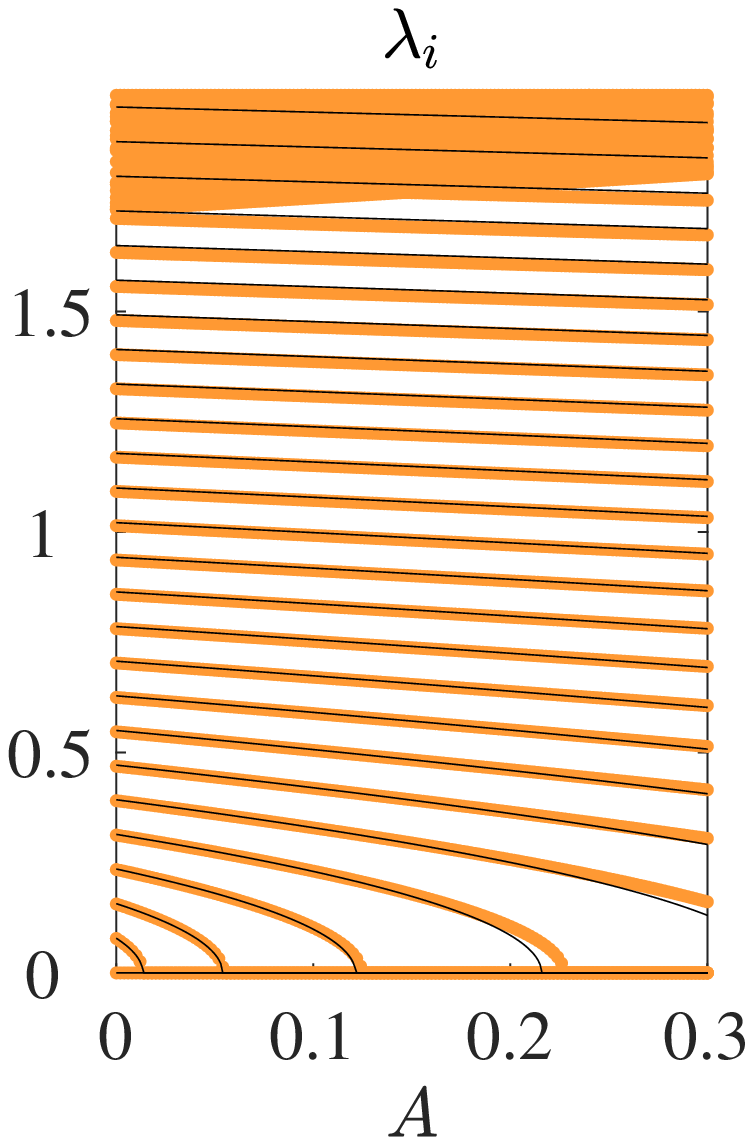}
\includegraphics[height=4.10cm]{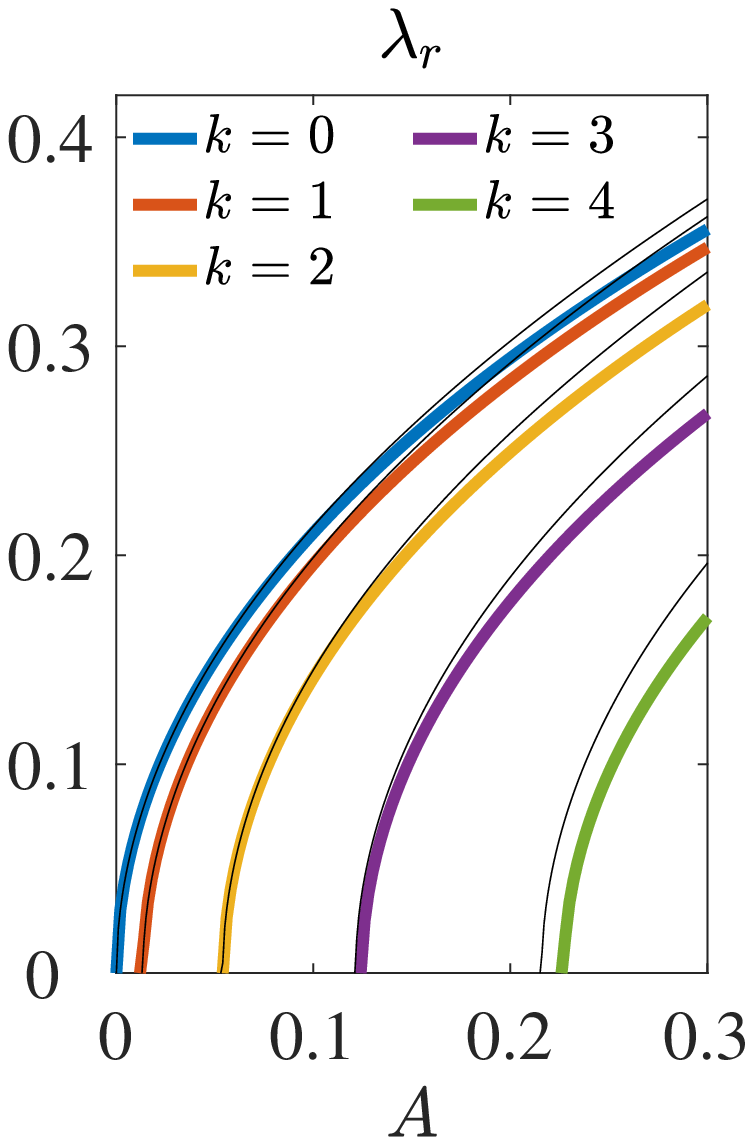}
\caption{(Color online)
Same as in Fig.~\ref{fig:evals_sG} but for the $\phi^4$ model.
Theoretical predictions are given by Eq.~(\ref{id_eq19}).
}
\label{fig:evals_phi4s}
\end{center}
\end{figure}

\begin{figure}[tb]
\begin{center}
\includegraphics[height=2.30cm]{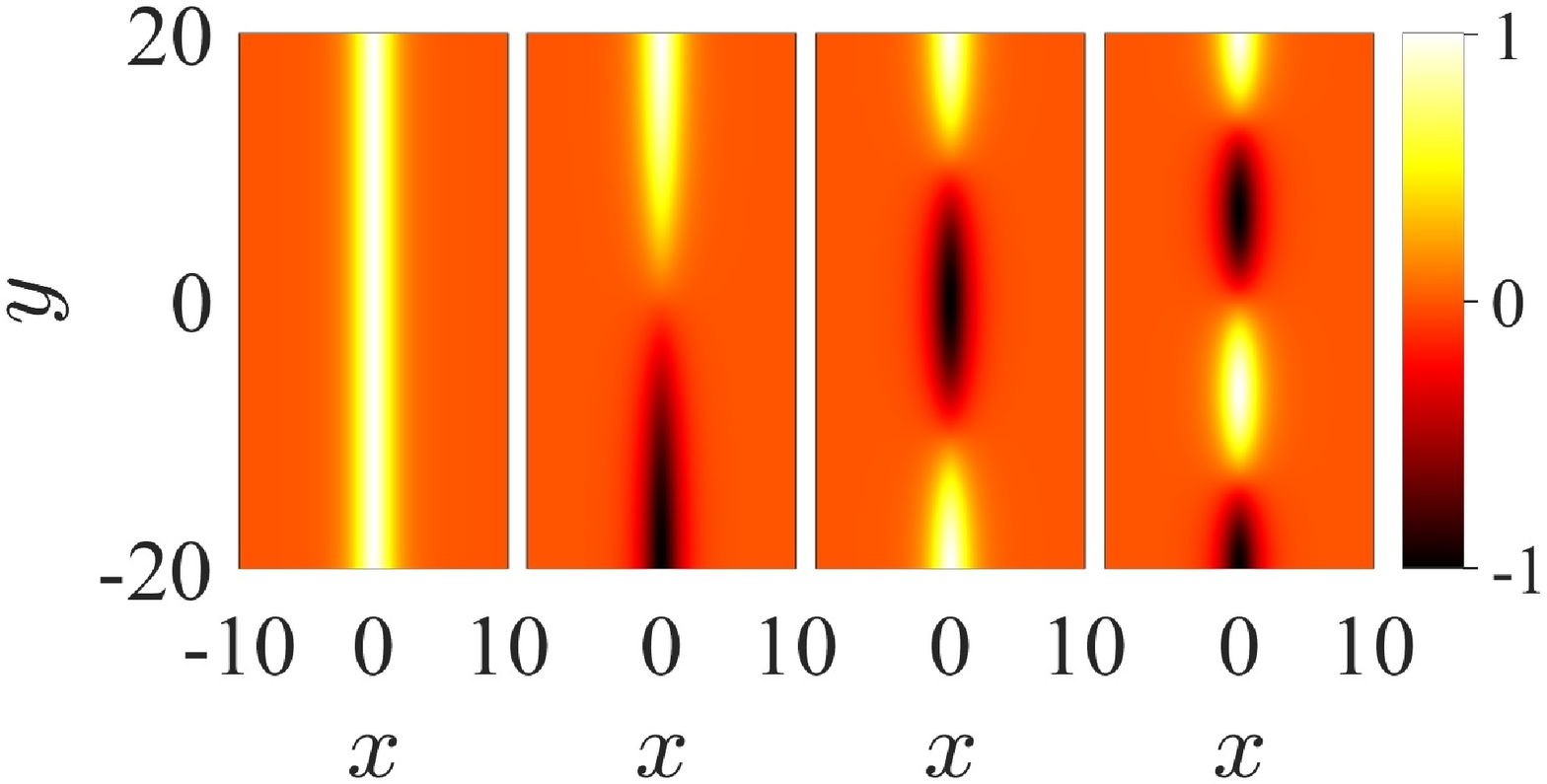}
~
\includegraphics[height=2.30cm]{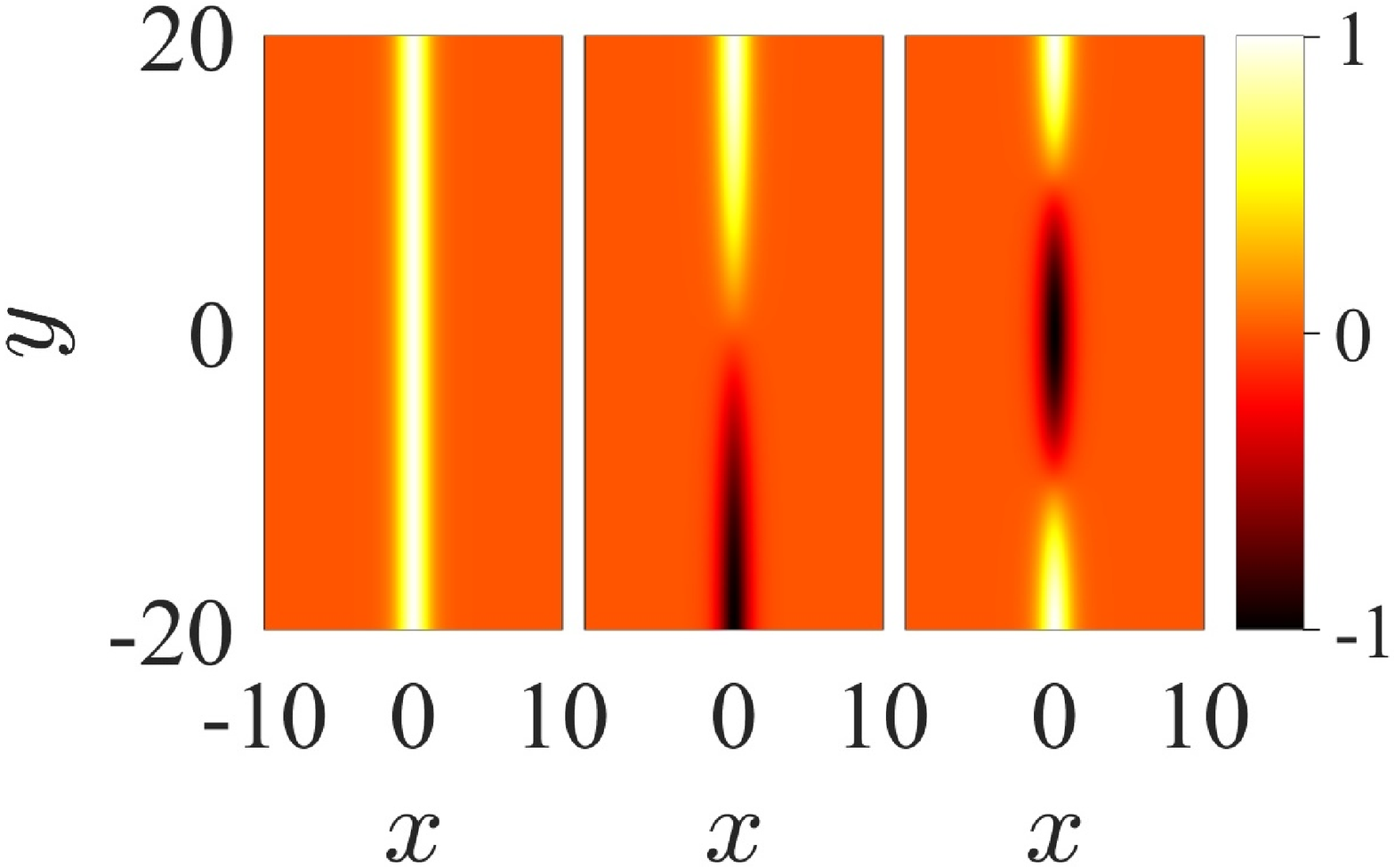}
\caption{(Color online)
Unstable eigenfunctions for the sG (left group of
panels) and the $\phi^4$ (right group of panels) models
corresponding to the steady states shown in Figs.~\ref{id_fig3} 
and \ref{id_fig4} in the presence of the longitudinal 
potential $V_{\rm ext}(x)=A\, {\rm sech}^2(x)$ for $A=0.25$ and
$A=0.1$, respectively.
Note that for these parameter values, the steady state for
the sG model possesses four unstable eigenvalues
($k=0$, $k=1$, $k=2$, and $k=3$) while for the steady
state for the $\phi^4$ model has three unstable eigenvalues
($k=0$, $k=1$, and $k=2$).
}
\label{fig:evecs}
\end{center}
\end{figure}

To get a better sense of the theory's capability at predicting the 
stability of the corresponding kinks in the presence of external potentials,
we depict in Figs.~\ref{fig:evals_sG} and \ref{fig:evals_phi4s} the linearization 
spectra as the amplitude $A$ of the longitudinal potential 
$V_{\rm ext}(x)=A\, {\rm sech}^2(x)$ is varied. 
As can be seen, overall, both in the absence of a
potential and in the presence of a longitudinal one, 
the theory yields the correct qualitative and a good quantitative
picture for the existence and stability of the kink.
This predisposes us to believe that the AI approach might also give
an accurate description of the full, nonlinear, dynamics of the kink
filament. It is important to note that the theory allows us to obtain
an immediate sense of whether the kink will be stable or unstable
(and how unstable, if it is indeed unstable).

Let us now focus on the dynamics of the kinks. As detailed above, in the 
presence of the external potential $V_{\rm ext}(x)=A\, {\rm sech}^2(x)$,
the stationary kink becomes immediately unstable for $A>0$. The dynamics
of this instability will be mediated by the presence of eigenfunctions
associated with unstable eigenvalues.
In Fig.~\ref{fig:evecs} we plot the unstable eigenfunctions
corresponding to the case presented in Figs.~\ref{id_fig3} and 
\ref{id_fig4}.
These unstable eigenfunctions will dictate the initial destabilization
of the steady states. Note that for both, the sG and the $\phi^4$
cases, the $k=0$ and $k=1$ modes have very similar corresponding
eigenvalues. Therefore, we expect the destabilization dynamics to
follow, predominantly, a combination of these two modes.
%
%
Note that given the
form of the perturbation expansion in Eq.~(\ref{eq:perturb}), 
the eigenfunctions should also
be used to construct suitable perturbations of the velocity field at $t=0$
according to $u_t(x,y,t)=\epsilon \lambda e^{\lambda t} w(x,y)$.
Thus, in Fig.~\ref{fig:evecs} a light region indicates
movement to the right while a dark region indicates movement to the left
(or vice versa).
As such, the first unstable ($k=0$) mode for each model corresponds to a
{\em translational} mode that destabilizes the kink and makes it
go ``down the hill'' from the external potential.
The next mode, associated with $k=1$, will cause the kink to snake 
such that half of it goes to the right and the other half goes to the 
left of the external potential hill (or vice versa).
Similarly, other (higher) unstable modes will induce snaking of the
kink according to the wavenumber $k$.

\begin{figure}[tb]
\begin{center}
\includegraphics[height=4.01cm]{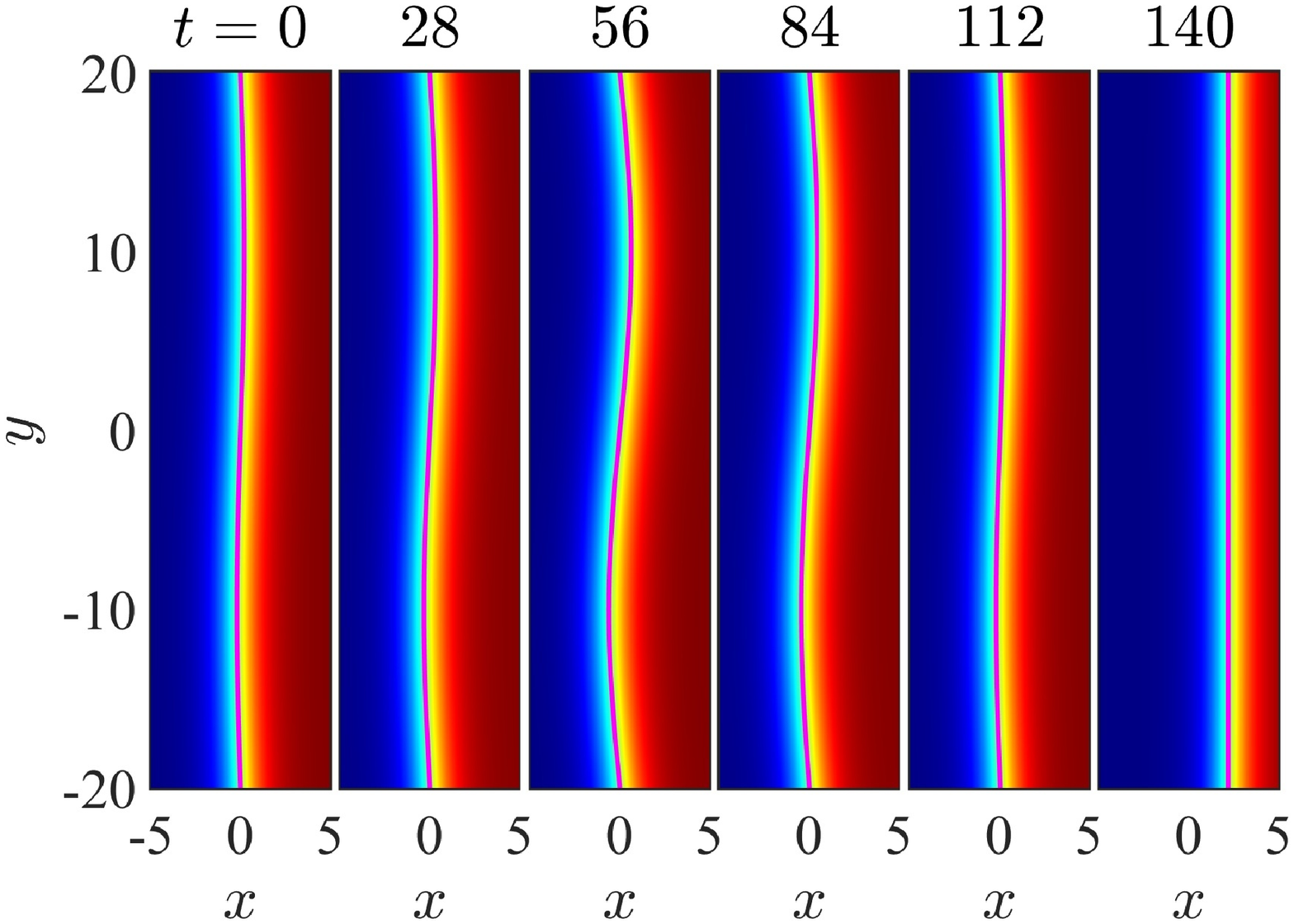}
\\[2.0ex]
\includegraphics[height=4.01cm]{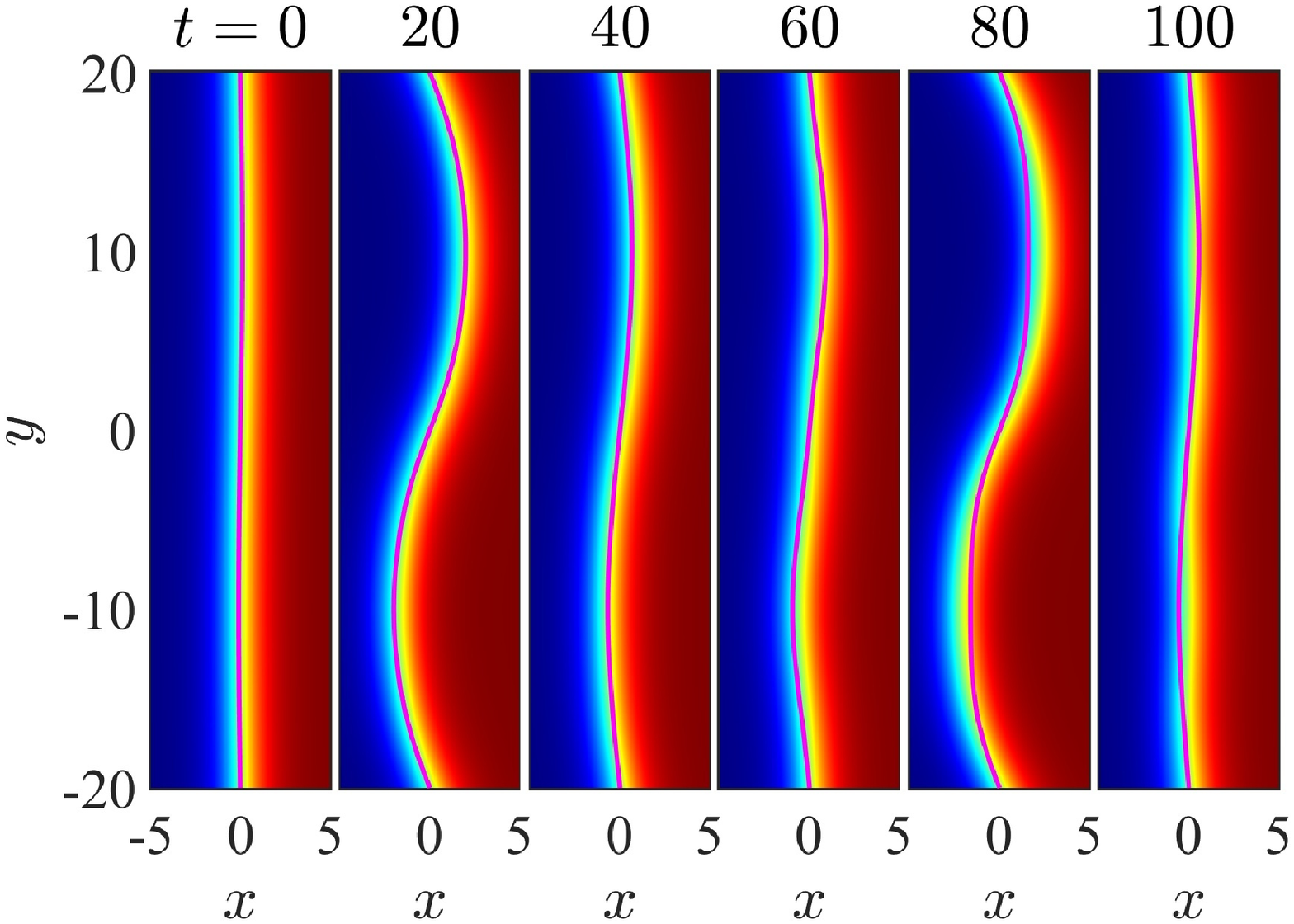}
\caption{(Color online)
Dynamics of a sG kink under the longitudinal potential
$V_{\rm ext}(x)=A\, {\rm sech}^2(x)$ for $A=0.1$ (top panels)
and $A=0.25$ (bottom panels) at the indicated times.
The background image depicts the field $u(x,y,t)$ while the magenta line
depicts the corresponding dynamics using the reduced AI 
PDE~(\ref{id_eq14}) with $P'(X)$ defined in Eq.~(\ref{id_eq16}).
Both systems are initialized with the same initial condition
corresponding to a stationary kink perturbed with the $k=2$
longitudinal mode with amplitude 0.1. Namely, the initial location 
of the kink is given by $X(y,t=0)=\varepsilon\,\sin(\pi ky/L_y)$
with $\varepsilon=0.1$ and $k=2$.
See Supplemental Material {\tt movie-sG-1} and {\tt movie-sG-2} 
for animations depicting the corresponding dynamics~\cite{SupMat}.
%
%
}
\label{fig:AI_vs_sG1}
\end{center}
\end{figure}

\begin{figure}[tb]
\begin{center}
\includegraphics[height=4.01cm]{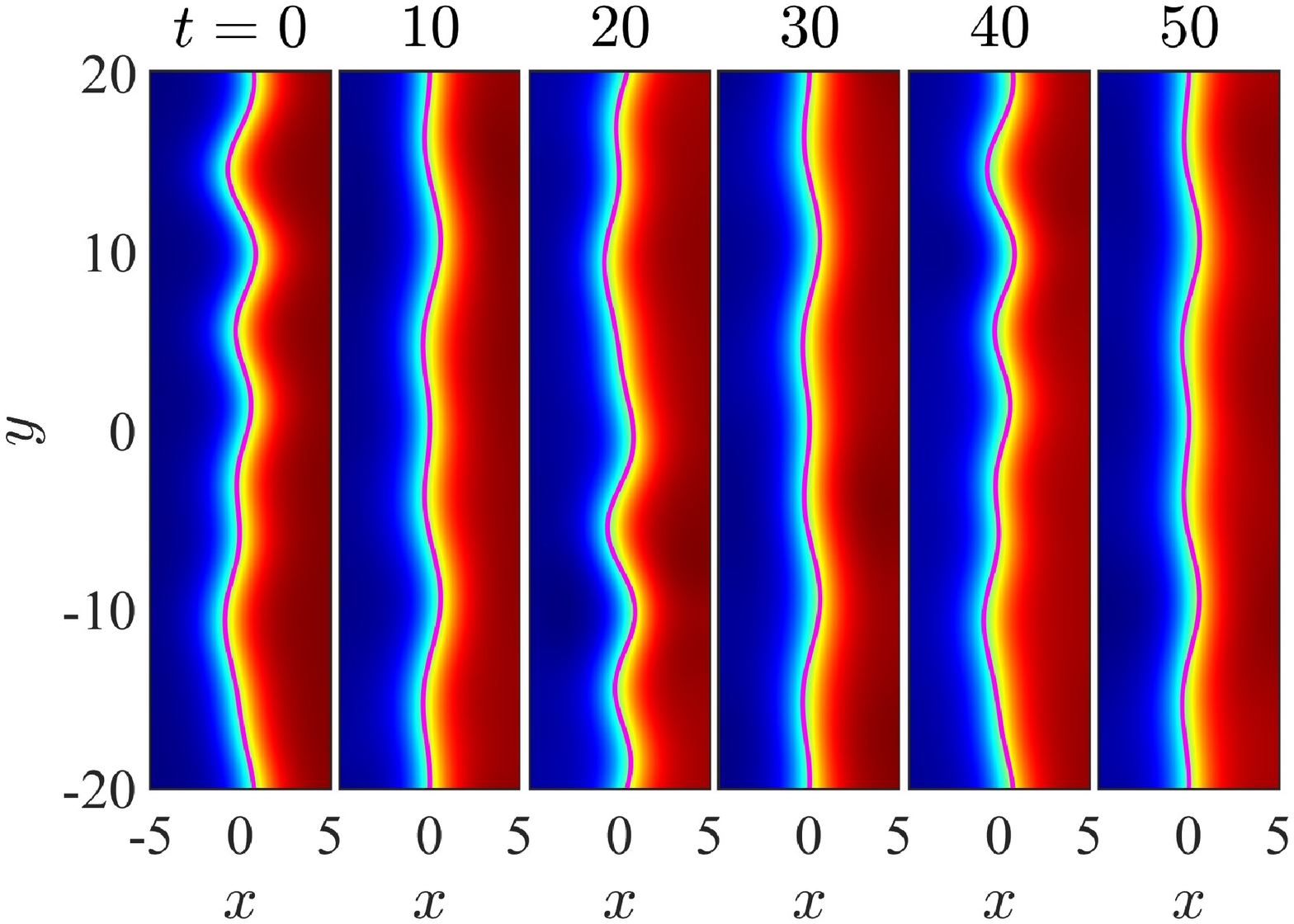}
\\[2.0ex]
\includegraphics[height=4.01cm]{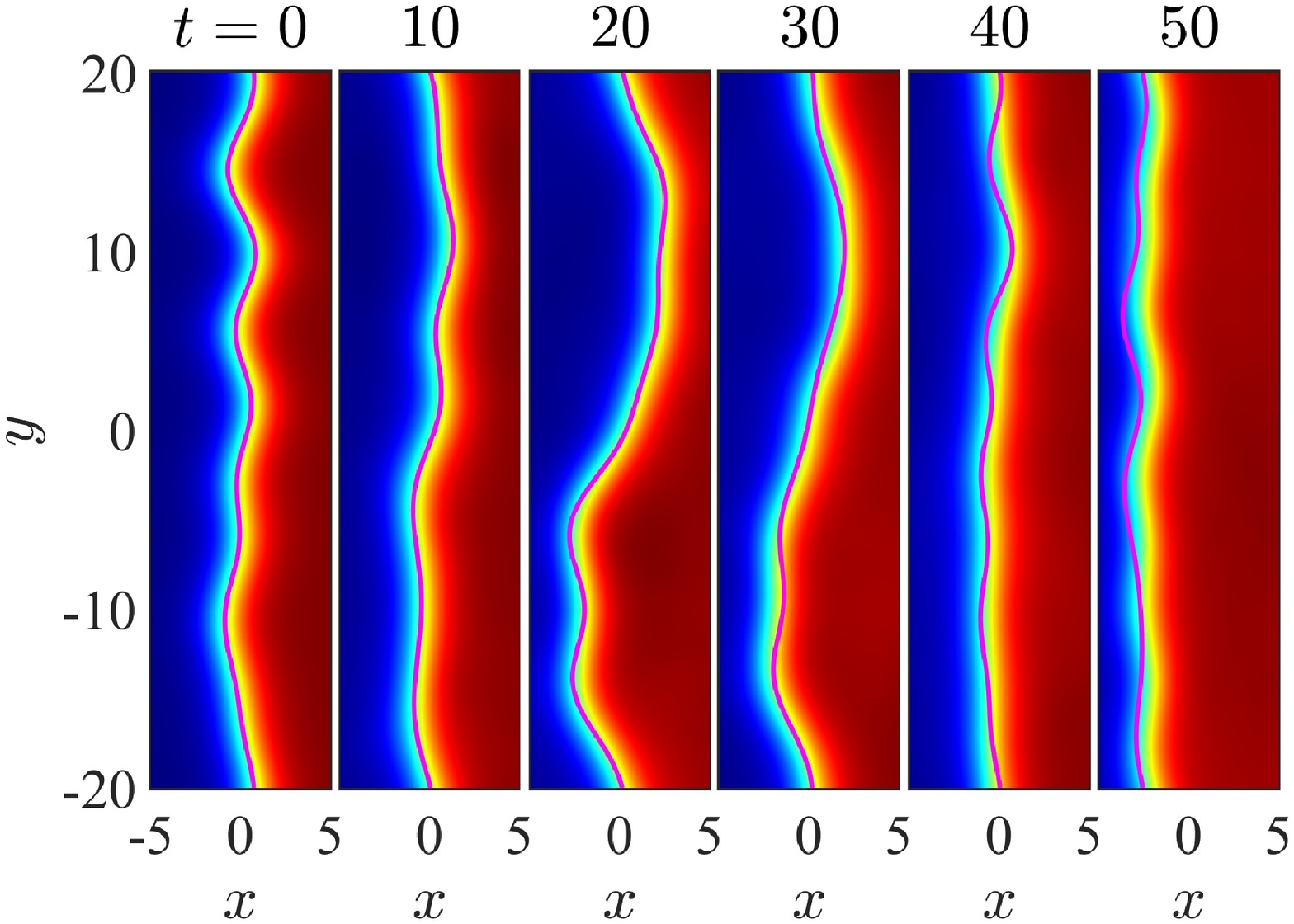}
\caption{(Color online)
Same as in Fig.~\ref{fig:AI_vs_sG1} but for $A=0$ (top panels)
and $A=0.25$ (bottom panels) for an initial condition corresponding
to a stationary kink perturbed with a (linear) combination of the 
modes $k=2,4,6,8,10$.
Specifically, $X(y,t=0)=\sum_{j=1}^{5}\varepsilon_j\,\sin(2\pi jy/L_y+\varphi_j)$
with $\varepsilon_j=0.3$, $\varphi_j=(j-1)L_y\pi/10$, and $k=2j$.
See Supplemental Material {\tt movie-sG-3} and {\tt movie-sG-4} 
for animations depicting the corresponding dynamics~\cite{SupMat}.
}
\label{fig:AI_vs_sG2}
\end{center}
\end{figure}

\begin{figure}[tb]
\begin{center}
\includegraphics[height=4.01cm]{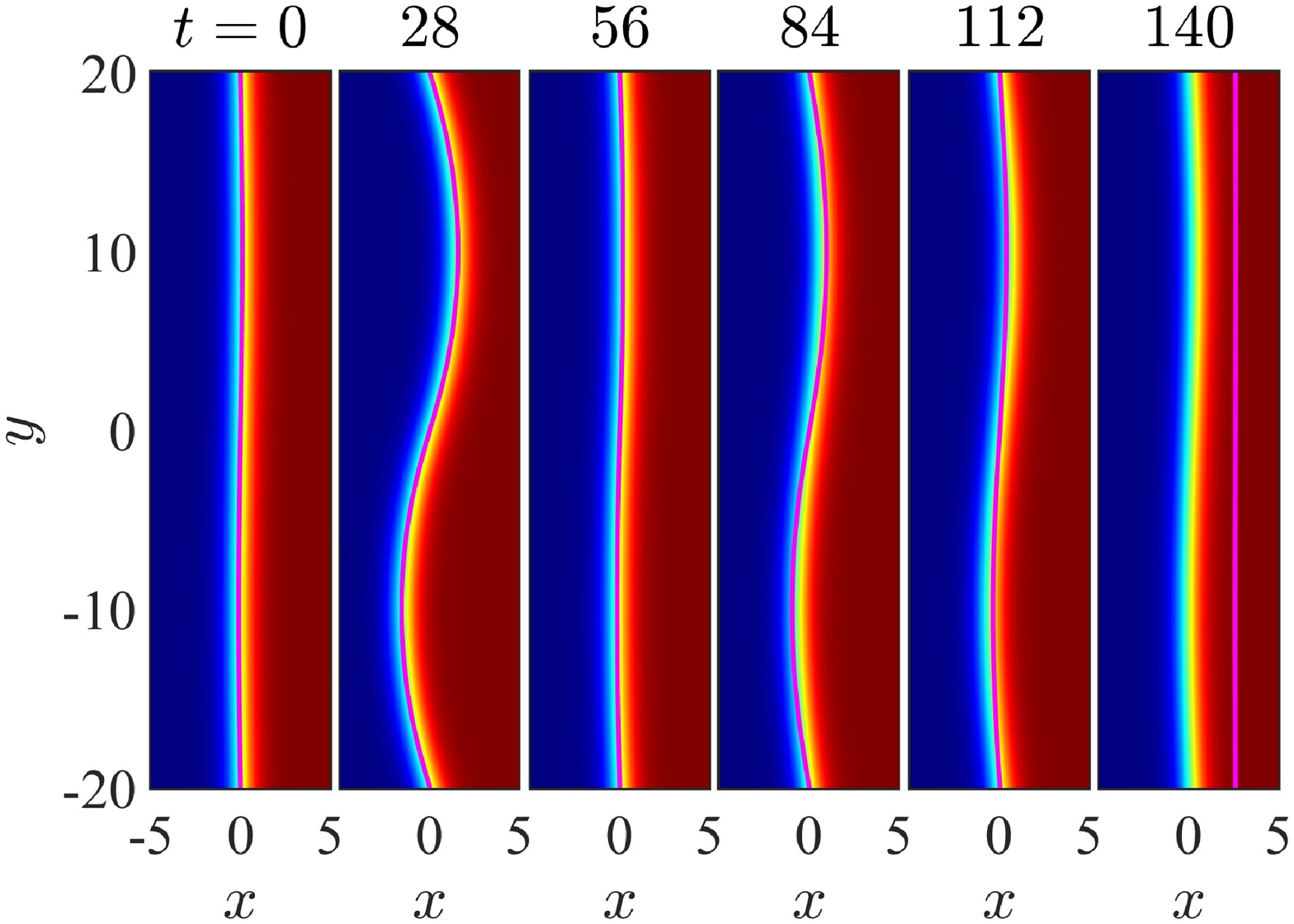}
\\[2.0ex]
\includegraphics[height=4.01cm]{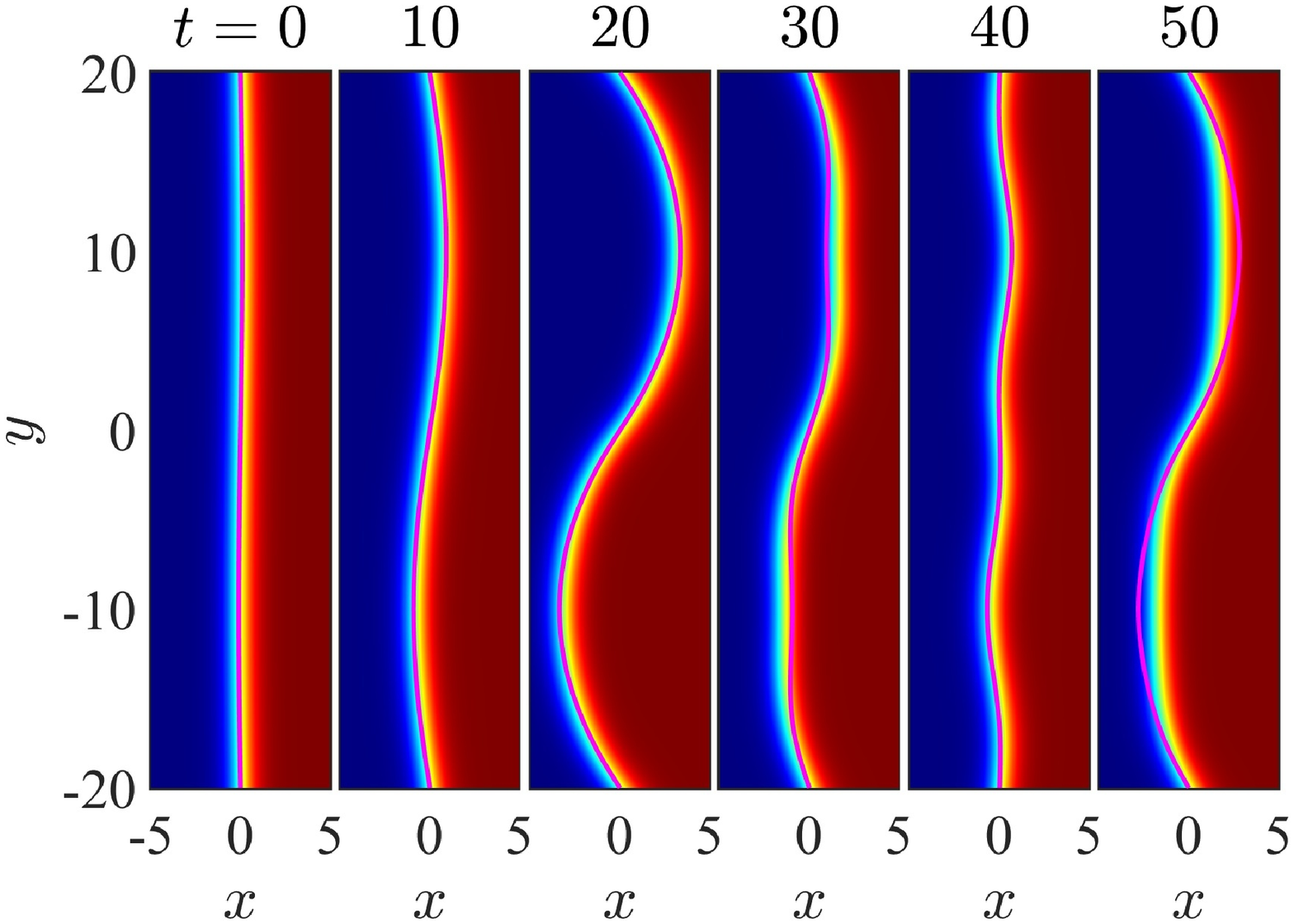}
\caption{(Color online)
Same as in Fig.~\ref{fig:AI_vs_sG1} but for the $\phi^4$ model.
See Supplemental Material {\tt movie-phi4-1} and {\tt movie-phi4-2} 
for animations depicting the corresponding dynamics~\cite{SupMat}.
}
\label{fig:AI_vs_phi41}
\end{center}
\end{figure}

\begin{figure}[tb]
\begin{center}
\includegraphics[height=4.01cm]{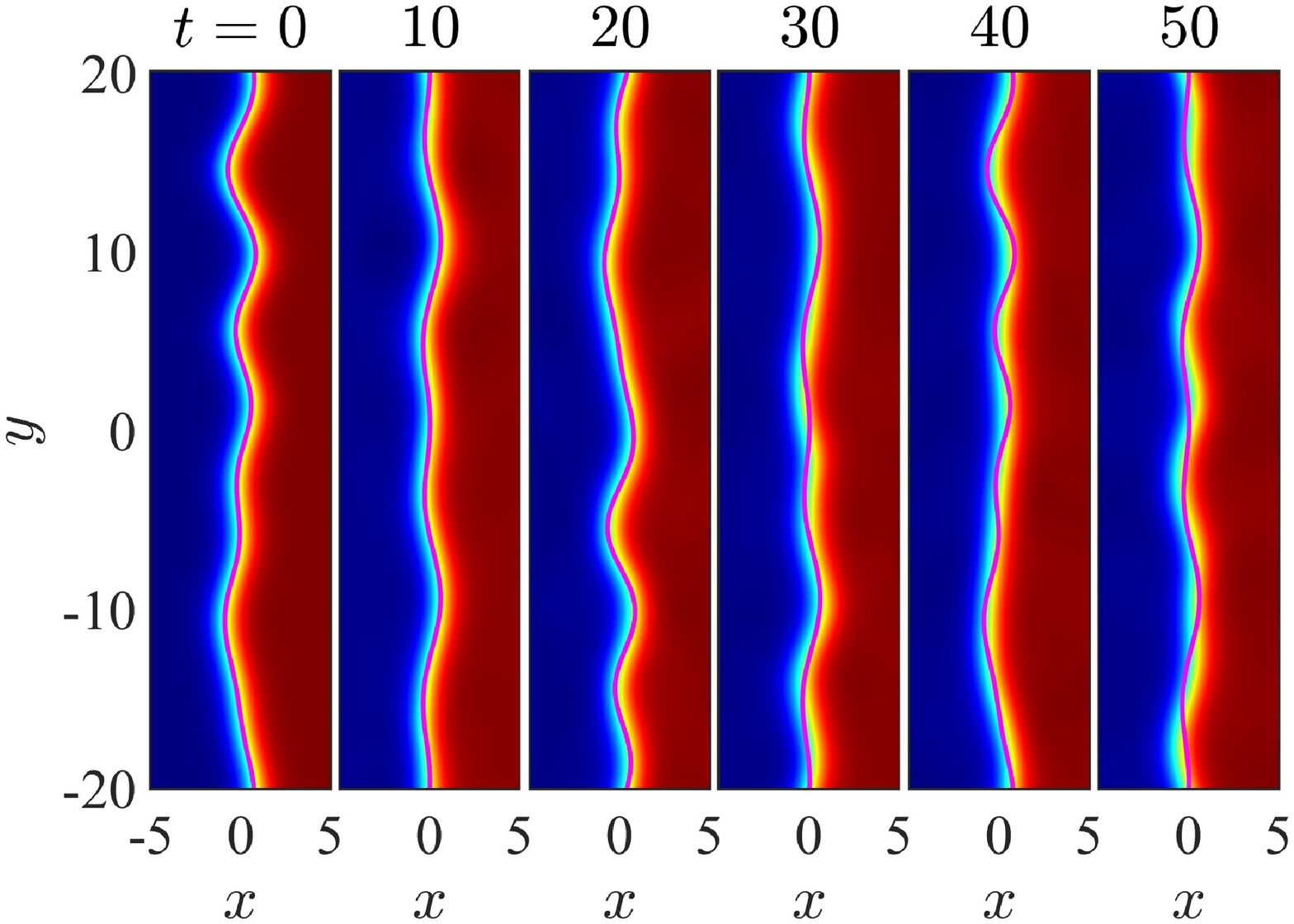}
\\[2.0ex]
\includegraphics[height=4.01cm]{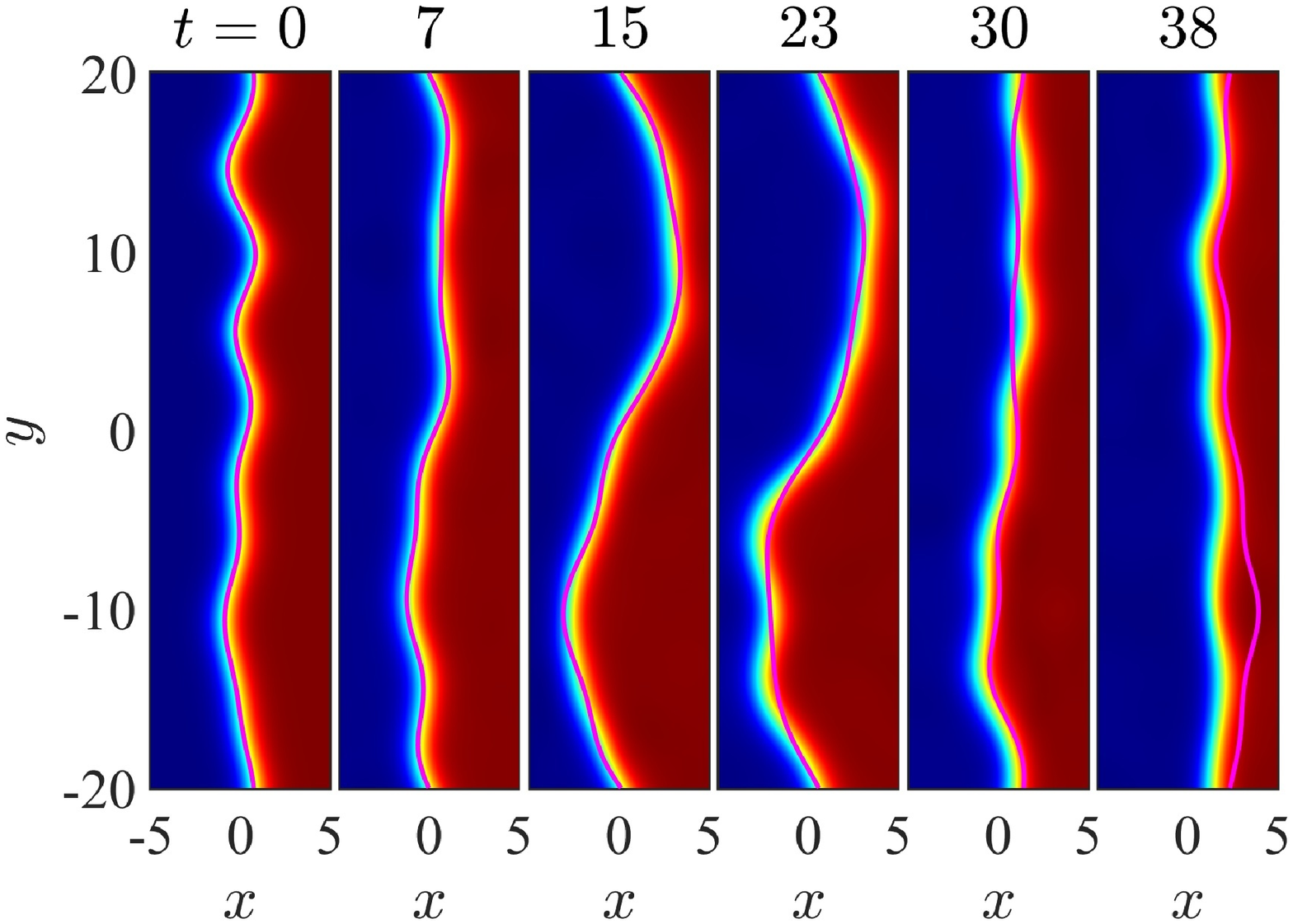}
\caption{(Color online)
Same as in Fig.~\ref{fig:AI_vs_sG2} but for the $\phi^4$ model.
See Supplemental Material {\tt movie-phi4-3} and {\tt movie-phi4-4} 
for animations depicting the corresponding dynamics~\cite{SupMat}.
}
\label{fig:AI_vs_phi42}
\end{center}
\end{figure}

So far, we have shown that the AI does a very good job at predicting the linearized 
behavior around the steady state. Therefore, at this stage, we would like to 
directly compare the full (nonlinear) dynamics of the sG and $\phi^4$
 kinks to that predicted by our AI reduction.
Figure~\ref{fig:AI_vs_sG1} shows a comparison, for a couple of cases,
between the full sG dynamics and the reduced AI PDE~(\ref{id_eq14}) 
with $P'(X)$ defined in Eq.~(\ref{id_eq16}).
As can be observed, the AI reduction is able to predict the
full nonlinear dynamics of the kink. In particular in these two cases, 
the dynamics is a combination of the oscillations of the $k=2$ mode
and the destabilization through the translational $k=0$ mode. 
Further numerical experiments (not shown here) show that this combined
dynamics is rather general for a wide range of initial conditions of the 
kink of the form $X(y,t=0)=\varepsilon\,\sin(\pi ky/L_y)$ 
for $k>0$ (or combinations thereof for different values of $k$).
Namely, the typical dynamics is one where the different $k$ modes are 
initially destabilized according to their unstable eigenvalue (see
the linear stability results above) and as they grow they enter the nonlinear
regime of the dynamics 
and appear to return to the vicinity of the initial stationary kink. 
At the same time, perturbations along the $k=0$
(translational) mode ---seeded by the initial condition or numerically
seeded by the finite precision of the numerics--- push the kink 
completely to one side of the crest of the longitudinal external potential.
Then the kink continues to oscillate along the perturbed $k>0$ modes
while it has gained horizontal (in the $x$-direction) speed and keeps
traveling towards the boundary of the domain.
In Fig.~\ref{fig:AI_vs_sG2} we further test the capability of
the reduced AI PDE in predicting the full sG dynamics by initializing 
the kink with a perturbation that includes a (linear) combination of
the modes $k=2,4,6,8,10$ in the absence (top panels) and
presence (bottom panels) of the longitudinal potential.
In the absence of an external potential the kink is (neutrally) stable, and
as such it only ``wiggles'' in a linear fashion
[i.e., each of the $k$ modes oscillates according to its frequency
given by Eq.~(\ref{id_eq18})].
Perhaps more interesting is the dynamics under the influence of the
longitudinal potential as the lowest-lying modes (in this case $k=0,1,
2,3$) are unstable. In this case, as seen in the bottom panels of 
Fig.~\ref{fig:AI_vs_sG2}, the dynamics is more involved as it is
fully nonlinear. Nevertheless, the AI reduction is able to closely
emulate the dynamics of the full system even in this fully
nonlinear regime.

Finally, in Fig.~\ref{fig:AI_vs_phi41} and \ref{fig:AI_vs_phi42} we present
the equivalent results to those presented in Figs.~\ref{fig:AI_vs_sG1} and
\ref{fig:AI_vs_sG2} but for  $\phi^4$.
As it could be anticipated, the AI reduced model is also able to predict
the full $\phi^4$ dynamics not only in the linear regime but also in the 
fully nonlinear regime for considerably long times.
Nevertheless, the slight deviations identified in the instability growth
rates will eventually affect the quantitative matching between the two
for sufficiently long time scales as can be seen in these figures.
%

\subsection{Radial KG Kinks}

\begin{figure}[tb]
\begin{center}
  \includegraphics[height=9.0cm]{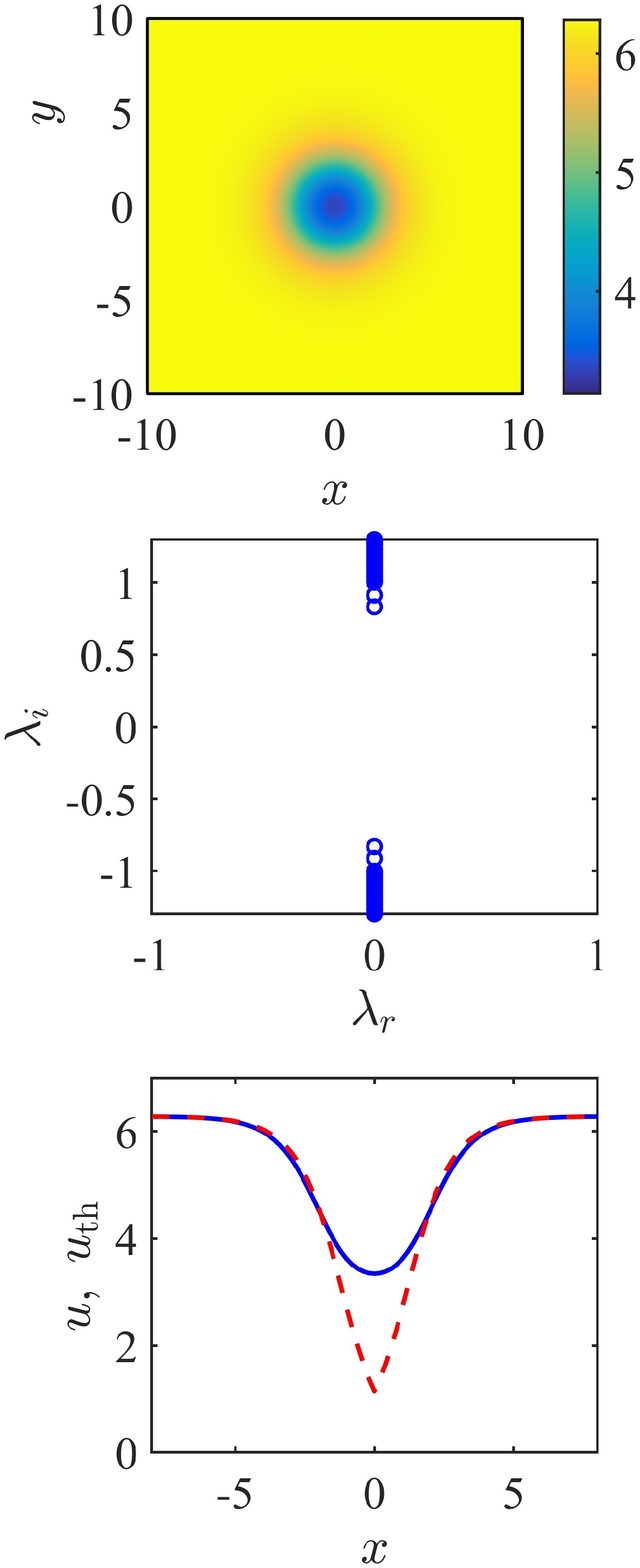}
  \includegraphics[height=9.0cm]{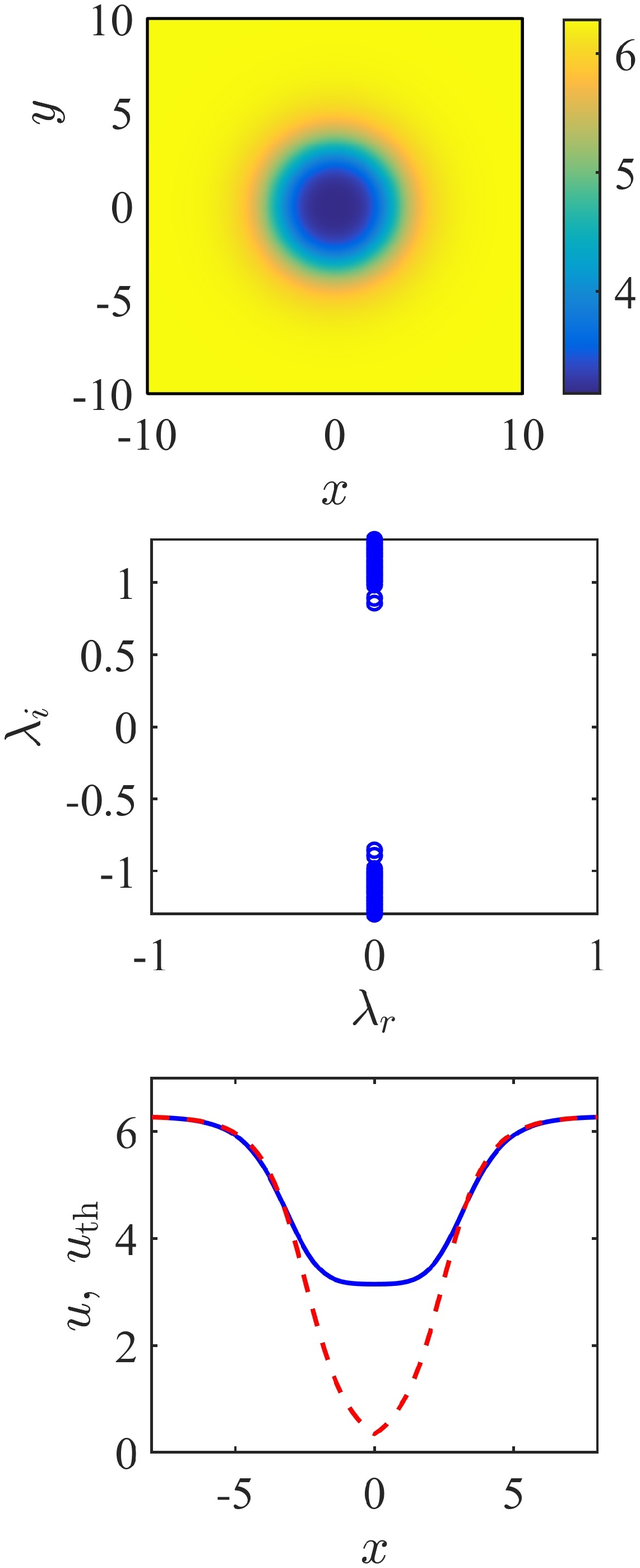}
  \caption{(Color online)
Stationary radial kink for the sG model under and external potential
    of the form $V_{\rm ext}(r)=A\, {\rm sech}(r)$. The top two
    sets of panels correspond to $A=-6$ (left)
    and $A=-16$ (right). The top panels show the two-dimensional
    radial kink solution, while the middle panels confirm that these solutions
    are dynamically stable by showing their spectrum (residing on the
    imaginary axis). The bottom panels compare the exact
    numerical solution (solid blue line) with the theoretical prediction
    $u_{\rm th}$ (red dashed line) for such an equilibrium given
    by a sG kink centered at $R_0$ given by solving Eq.~(\ref{eq:R0}). 
    The corresponding theoretical values
    for the equilibrium radius are $R_0=1.2212$ for $A=-6$
    and $R_0=2.4541$ for $A=-16$.
}
\label{id_fig5}
\end{center}
\end{figure}

\begin{figure}[tb]
\begin{center}
  \includegraphics[height=9.0cm]{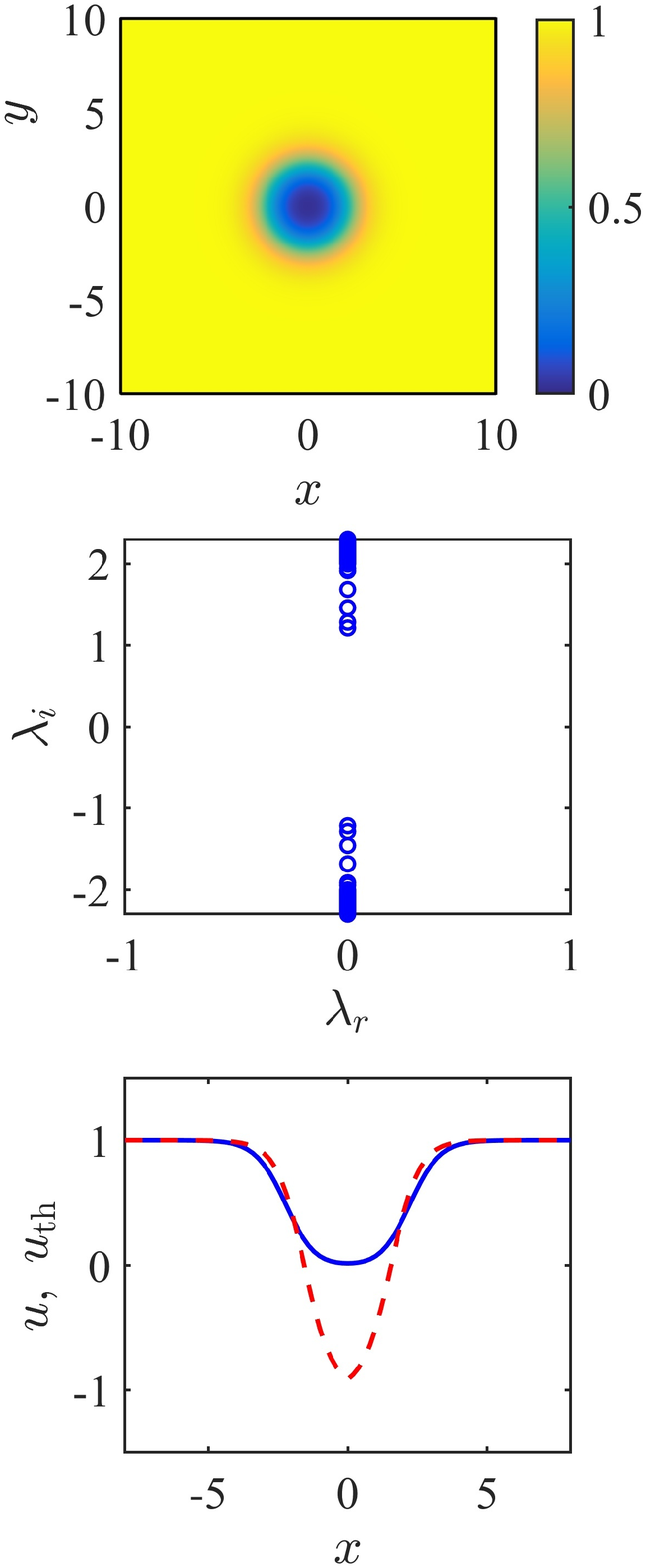}
  \includegraphics[height=9.0cm]{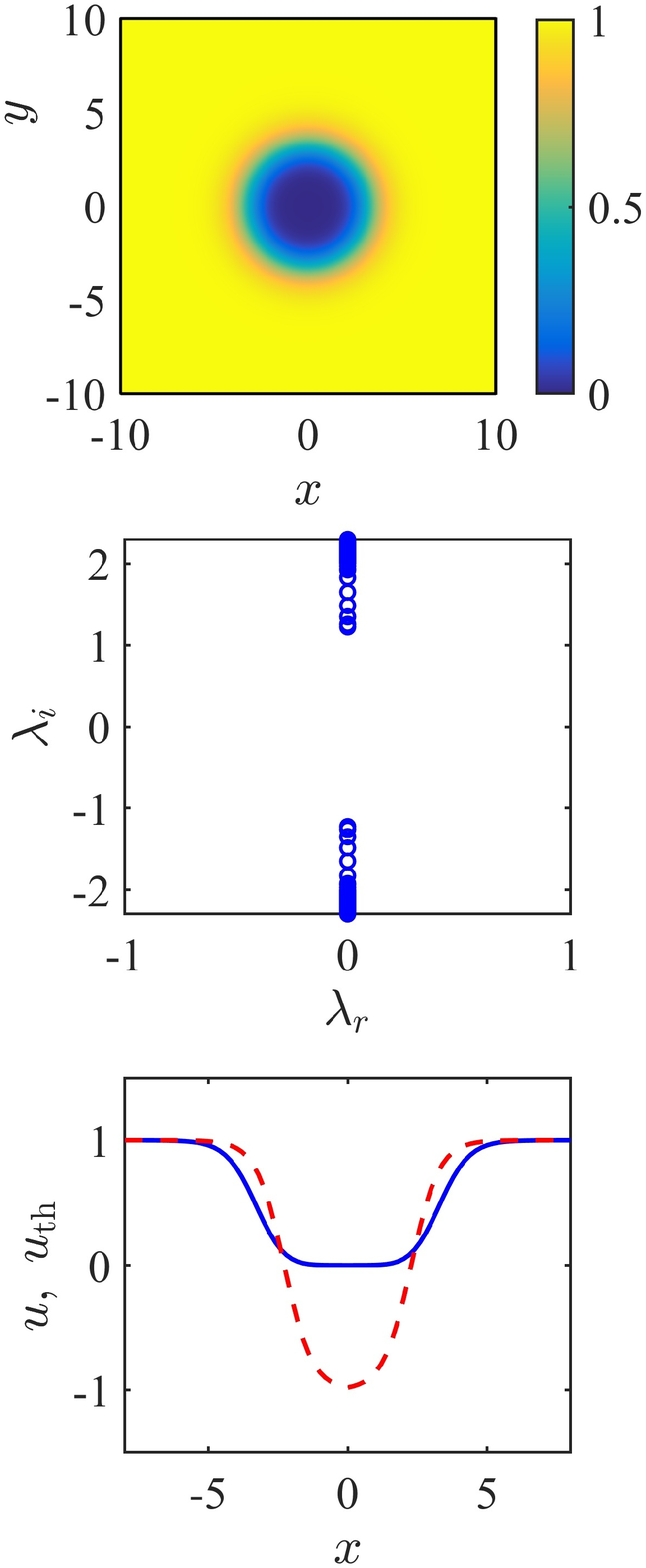}
  \caption{(Color online)
  Exactly the same diagnostics as for the
  sG equation and for the same potential
  but now for the case of
  the radial potential in the $\phi^4$ model.
  The corresponding theoretical values for the equilibrium 
  radius from Eq.~(\ref{eq:R0}) are $R_0=1.5565$ for $A=-6$
  and $R_0=2.2812$ for $A=-16$.
}
\label{id_fig6}
\end{center}
\end{figure}

We now turn to the case of the radial kinks. Here
the theory provides a very useful guideline about inducing
an unprecedented (to the best of our knowledge) steady state radial
kink. Namely, the analytical results serve to guide the intuition of
how to select an effective potential that counters the inward
force exerted on the kink by curvature. In so doing, this ring
potential succeeds in stabilizing the kink against this inward
collapse and allows it to execute {\em stable} oscillations
around the selected
equilibrium position. Moreover, the theory serves to explain the fact
that the azimuthal undulations do not destabilize the kink but simply
correspond to stable oscillatory modes (akin to Kelvin
waves in the BEC realm~\cite{siambook}). Indeed, everything once again
seems to be entirely the opposite of the defocusing NLS case (confirming
a similarity to a focusing rather than a defocusing nonlinearity).
In particular, in repulsive atomic BECs described by defocusing NLS,
the curvature pushes the dark solitons outward, while the trap induces
them a restoring force enabling the equilibrium. Around this equilibrium,
the undulations are unstable, leading to the formation of
vortices~\cite{djf,siambook}. For the KG case however,
the inward effect of curvature is
countered, as is shown in Figs.~\ref{id_fig5} (for sG) and~\ref{id_fig6}
(for $\phi^4$), by a suitable external potential. In the latter setting
the undulations are purely oscillatory and the radial state is spectrally
stable.

Using the information provided by the AI analysis of Sec.~\ref{sec:radial},
we searched for a stationary solution of the radial Klein-Gordon equation
\begin{equation} 
\label{rad_kg}
u_{tt} = u_{rr} + \frac{1}{r} u_{r} - (1+ V_{\rm ext})V'(u)=0,
\end{equation}
using Newton iterations. Explicitly, the Newton iteration to obtain
the next iterate $u^{n+1}$ in terms of the current iterate $u^n$
can be cast as
\begin{eqnarray} 
&&\left[\partial_{rr}+\frac{\partial_r}{r} - (1+ V_{\rm ext})V''(u^n)\right]u^{n+1}
\nonumber
\\[1.5ex]
\nonumber
&& \qquad\qquad = (1+ V_{\rm ext})\left[V'(u^n) -V''(u^n)\,u^{n}\right],
\end{eqnarray}
where $u^0$ is a small radial Gaussian initial guess. The
problem is discretized using finite differences where at $r=0$
we use l'H\^ospital's rule to regularize the term $u_{r}/r$.
As may be expected we obtain no steady states for $A>0$.
However, for $A<0$, we obtain rapid convergence to the profiles 
shown in the top panels of Figs.~\ref{id_fig5} and \ref{id_fig6}. As $A$ is
decreased, the solution at $r=0$ tends to $\pi$ for the sine-Gordon equation
($0$ for $\phi^4$) where it asymptotes. The central flat region where $u\approx\pi$
then increases its extent as $A$ decreases further. 

In the bottom panels in Figs.~\ref{id_fig5} and \ref{id_fig6} we compare the 
prediction (with a red dashed line) of the theoretical radial equilibrium
state of Sec.~\ref{sec:radial} against the full numerical finding of the corresponding
computation (in a blue solid line) for the potential $V_{\rm ext}$ given by
the (magenta) dash-dotted line. We can see that the theory only does a
moderately accurate job of capturing the ring equilibrium. However, a closer
inspection clarifies why this is rather natural to expect to be the case.
The equilibrium radius is rather small (i.e.,
between 1 and 2.5 in the cases
shown in the figures). In such a setting, the ansatz used is not
sufficiently accurate, as the approximations that we made in reaching
the filament PDEs are not valid.
In that sense, it is already quite encouraging that despite the
lack of validity of its assumptions, the theory does a fairly
reasonable job in capturing ---even if only in a sort of averaged
sense--- the rough profile and location of the actual steady state kink.
It is important to note in this
context that we also tried to examine cases of much stronger
potentials (cf., e.g., the top right panels in each of Figs.~\ref{id_fig5}
and~\ref{id_fig6}). In this case, an intriguing phenomenon arises
that merits further study. In particular, indeed the kink widens
as is expected from the theory since the force stemming from the
potential (which stabilizes against the inward curvature induced
motion) increases. However, instead of reaching all the way to $0$
for $r\approx 0$,
as a 1D kink would, the kink widens towards $\pi$ and then flattens
there (this is for the sG case ---an analogous feature happens for
$\phi^4$ with the $u=0$ state). It is remarkable that the saddle
point presents a form of ``impenetrable barrier'' and the kink
ends up forming in the radial setting between $\pi$ and $2 \pi$
for sG and between $0$ and $1$ for $\phi^4$. Once again, this warrants
further investigation and the potential use of a suitably adapted ansatz
to this setting.

For the purposes of the present work, we offer a qualitative
energetic argument about the existence of this state. At the level
of Eq.~(\ref{id_eq21}), for a stationary radial state, the first
(kinetic) and third (angular variation) terms in the energy are
absent. If, then, there is a connection between two different
steady states,
the homogeneous (i.e., independent of $V_{\rm ext}$)
fraction of the energy contributing to such
a state is given by 
\begin{eqnarray}
  E_{\rm rad}^{\rm 1K} = 2 \pi \int_0^{\infty} r \left[\frac{1}{2} u_r^2
    + V(u) \right] \,dr.
  \label{extra1}
\end{eqnarray}
In the limit of a potentially very thin kink centered at
$r=R$, this energy can be approximated as
$$
\nonumber
E_{\rm rad}^{\rm 1K }\approx 2 \pi R\,E_{\rm 1D}^{\rm 1K},
$$ 
where now $E_{\rm 1D}^{\rm 1K}$ is the quasi-1D
energy of this radial coherent structure.
However, this energetic contribution ``by itself''
is minimized at an $R=0$ radius, i.e., there is no term to balance it
and hence no such kink can arise without the presence of
an external potential term.
On the other hand, the inner state is at the saddle value $u=u_s$
(where $u_s=\pi$ for the sG model and $u_s=0$ for the $\phi^4$ model),
with positive energy $V(u_s) > 0$ [$V(u_s)=2$ for the 
sG model and $V(u_s)=1/2$ for the $\phi^4$ model]. 
Furthermore, if $V_{\rm ext}$ is non-zero (and, more specifically,
negative to offer a balancing energetic contribution), then
there also exists a ``bulk'' energetic contribution of the form:
\begin{eqnarray}
  E_{\rm rad}^{\rm ext} \approx 2 \pi \int_0^{\infty} r V_{\rm ext}(r) V(u) \,dr.
  \label{extra2}
\end{eqnarray}
In the limit of a thin filament of radius $r=R$, and if we have defined the
$V(u=0)=V(2 \pi)=0$ for sG [similarly $V(\pm 1)=0$ for $\phi^4$],
this energy can be approximated as:
\begin{eqnarray}
  E_{\rm rad}^{\rm ext}=2 \pi V(u_s) \int_0^{R} r V_{\rm ext}(r)  \,dr.
  \label{extra2a}
\end{eqnarray}
%
%
This becomes even simpler if, e.g., $V_{\rm ext}$ is a potential well
of depth $V_0$ in which case the integral simplifies further as 
%
$$
\nonumber
E_{\rm rad}^{\rm ext} \approx  -V(u_s) V_0 \pi R^2,
$$
which is clearly indeed a bulk contribution.
We can now add the approximate surface and bulk expressions obtained
above, take the extremum and obtain the approximate equilibrium radius 
$$
\nonumber
R_0 \approx \frac{E_{\rm 1D}^{\rm 1K}}{V(u_s) V_0}.
$$
%
%
This approximation is in the ballpark of our numerical findings and
offers some insight towards the energetic balance that gives rise
to the existence of this stable radial kink.

\begin{figure}[tb]
\begin{center}
  \includegraphics[width=4.3cm]{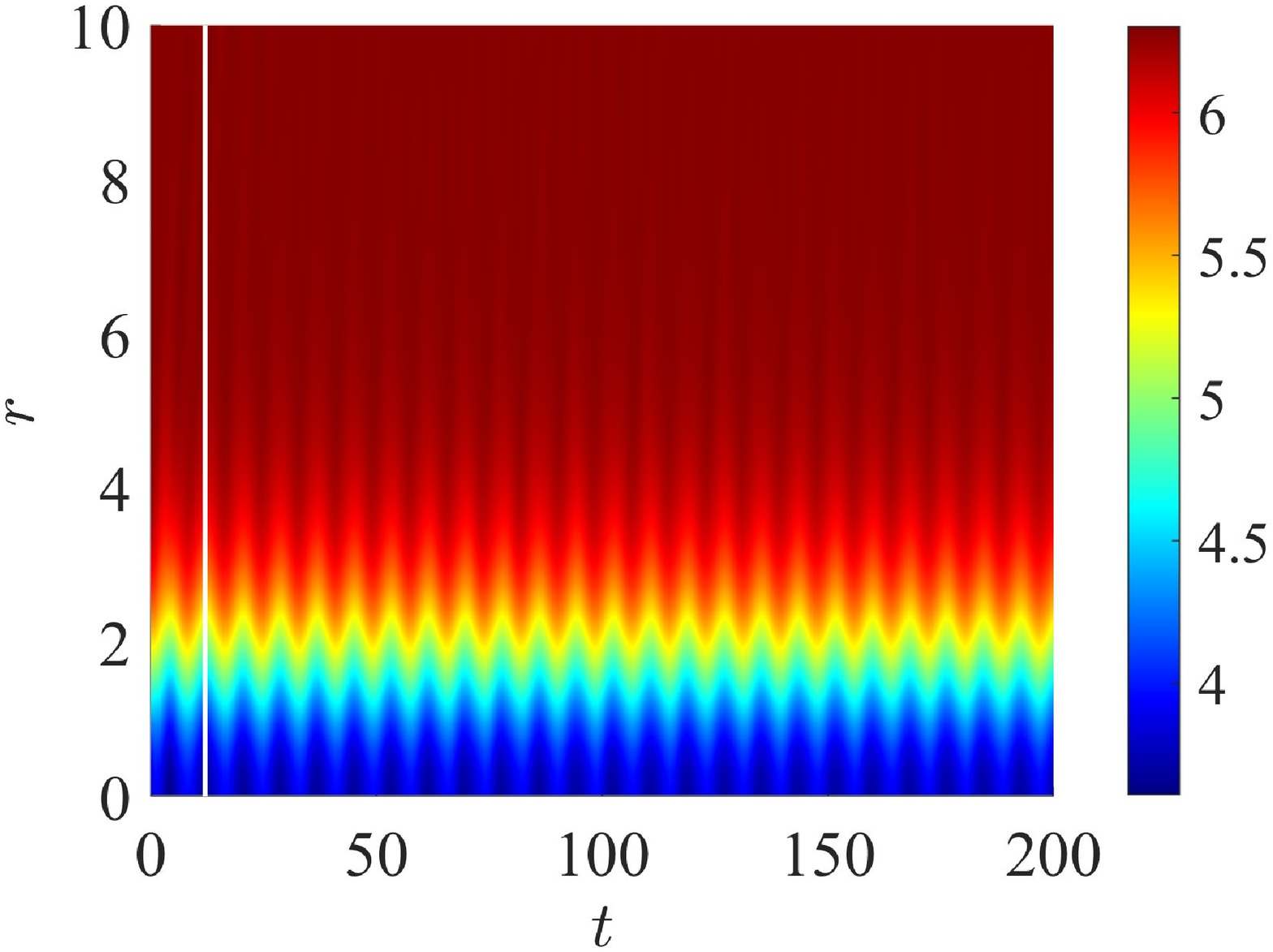}~
  \includegraphics[width=4.1cm]{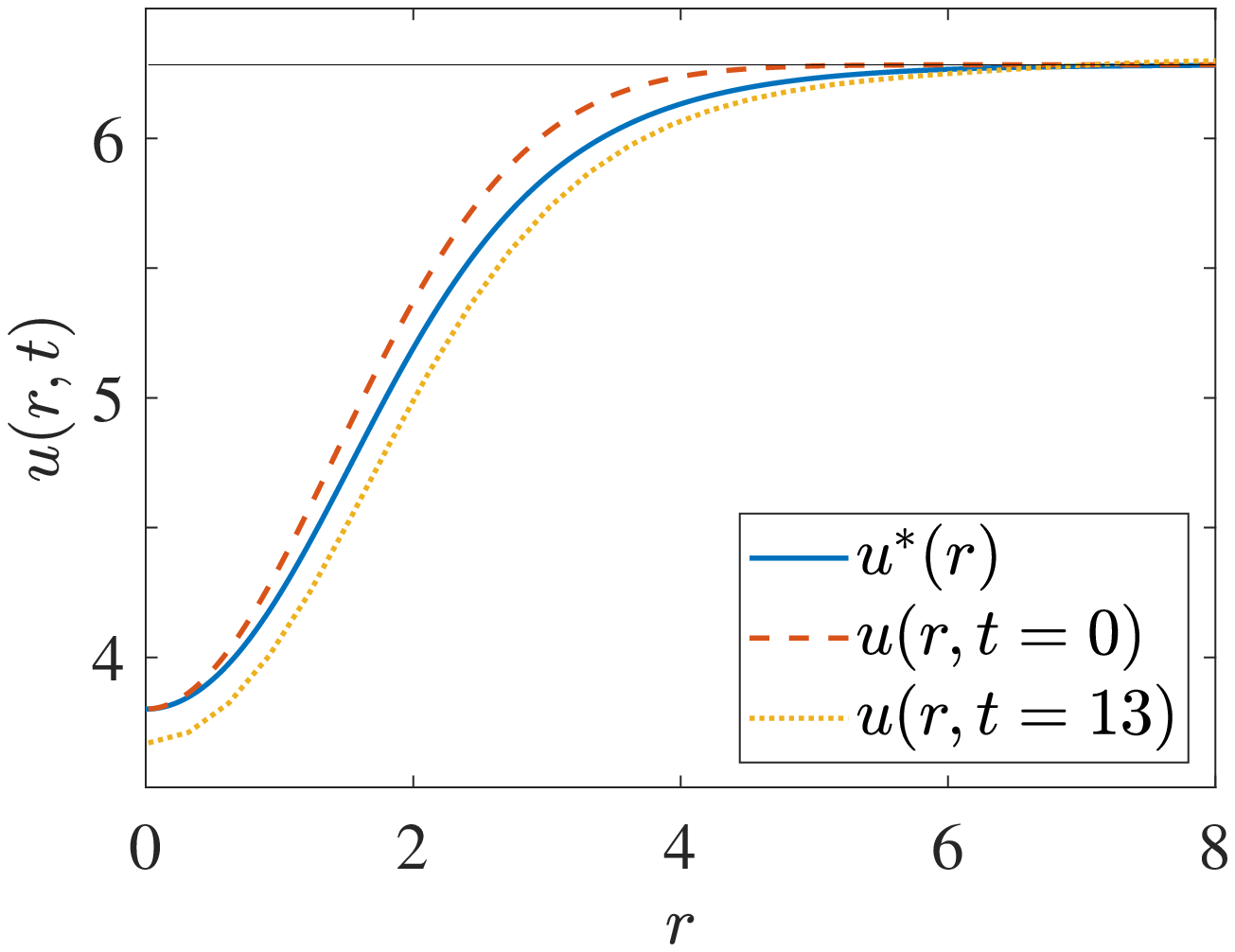}
  \caption{(Color online)
An example of the oscillatory radial evolution dynamics
    of the initial condition of Eq.~(\ref{extra3}) for the
    radial PDE of Eq.~(\ref{id_eq9}) and $V_{\rm ext}(r)=-4\, {\rm sech}(r)$.
    The left panel shows the $(r,t)$ space-time
    contour plot illustrating the stable
    vibrating dynamics of the radial kink. The right panel portrays
    individual cuts involving the state at $t=0$ (orange dashed)
    and the result at $t=13$ (yellow dotted; this time is
    denoted by a vertical line on the left panel). For comparison the exact
  stationary, stable radial configuration is depicted by a blue solid line.}
\label{id_fig7}
\end{center}
\end{figure}


The stability of this  radial kink was also tested dynamically
whereby we used the steady state kink (depicted by a blue solid  line 
in the right panel of Fig.~\ref{id_fig7}) as the initial condition of a 
time dependent radial sine-Gordon code (see Ref.~\cite{caputo} for
the details on the numerics). We observed no significant deviation
of the profile for times up to $t=3000$, indicating dynamical stability.
Furthermore, we also provided a larger perturbation of the kink
by initializing the system with the profile
\begin{eqnarray}
  u (r,0) = \pi (2 - 0.79) \exp\left(-\frac{r^2}{4}\right),  \quad 
  u_t(r,0)=0.
  \label{extra3}
\end{eqnarray}
depicted by a red dashed line in the right panel of Fig.~\ref{id_fig7}.
We see that despite the perturbed form of the initial condition,
the radial kink remains ``trapped'' in the effective potential well
of its radial energy landscape, oscillating robustly around the
stable minimum corresponding to the stationary, stable configuration
identified above. 
These results corroborate our spectral stability analysis.

\section{Conclusions \& Future Work}

In the present work, we have explored a variety of
settings related to the transverse dynamics of
kinks in KG models. We started from the realm
of planar (one-dimensional along, say, the $x$-direction)
kinks in a 2D domain. We illustrated the stability of
such structures against transverse undulations in the
realm of the adiabatic invariant theory of solitonic
filaments. Subsequently, we introduced external potentials
of different types (radial or longitudinal) and explored
under which conditions they could lead to stability
or immediate instability of the original kink structures.
Not only did we identify the qualitative conclusion regarding
this stability question; we also provided a systematic
set of predictions for the eigenvalues associated with
the transverse undulations of the kink. Even beyond that,
we have given an equation that describes the genuine nonlinear
dynamics of the kink as a solitonic filament embedded
within the 2D space.

We then turned to the case of radial kinks. So far, in this
setting the attempts have been to observe and characterize the
detrimental motion of the kink inward as induced by the $1/R$
effect of curvature. The most recent attempt was to utilize
this phenomenon as a source of fast breathers. The present work
moves one step further offering, on the basis of practically
accessible spatial inhomogeneities, the ability to produce
a force countering the curvature and producing a stationary
stable radial kinklike structure. 

Nevertheless, there are numerous challenges that the present
work raised towards future studies that are worthy of
further consideration. In the realm of longitudinal structures,
it would be interesting to explore whether not only
longitudinal potentials, but also
more general ones (such as those used in the right panels of
Figs.~\ref{id_fig1} and \ref{id_fig2} could be addressed within the
theory. In principle the theory does provide this possibility
at the level of the center of the structure considerations
given herein. However, the above figures suggest that perhaps
using ans{\"a}tze with further variables (such as the kink width)
may be more suitable to tackle such a setting.

A perhaps wider range of challenges awaits regarding the radial
kink case. Here, the adiabatic invariant formulation provides
a useful guideline but not a quantitative diagnostic. Extending
the theoretical considerations beyond the limitations and assumptions
detailed herein is a significant challenge that is certainly
worthy tackling. However, even at the purely numerical level
(and at the level of associated mathematical analysis) there
are surprises here. Perhaps the predominant one in this vein
is the feature identified in both the sG and $\phi^4$ models, 
whereby as the potential
strength is enlarged, a coherent structure is created
connecting the former saddle point (e.g. $u=\pi$ in sG or
$u=0$ in $\phi^4$) with the asymptotic value (of $u=2 \pi$ in
sG or $u=\pm 1$ in $\phi^4$). It is as if this saddle point
operates as an impenetrable barrier for the asymptotics of the state
in such a higher dimensional setting. This clearly merits some
theoretical understanding, further numerical exploration and
potentially a modified adiabatic invariant theory utilizing a suitable
structure for such an asymptotic state.

Additionally, one can envision numerous further generalizations,
including the consideration of general (rather than purely
radially symmetric) potentials in 2D, as well as the promising extension
of the present considerations in planar, spherical or more complex
3D patterns. Such studies are presently in progress and will be
reported in future publications. 

\acknowledgments

PGK and RCG gratefully acknowledge the kind hospitality of the University
of Rouen Normandy, Laboratoire de math\'ematiques Rapha\"el Salem,
where part of this research was conducted. This research is based
upon work supported by the National Science Foundation, under grants
PHY-1602994 and PHY-PHY-1603058.

\end{document}